\documentclass[oneside,12pt]{toddbook}
\usepackage{todd,afterpage,fancyheadings,cite,graphicx,threeparttable}

\def\lesssim{\mathrel{\hbox{\rlap{\hbox{\lower4pt\hbox{$\sim$}}}\hbox{$<$}}}}
\def\gtrsim{\mathrel{\hbox{\rlap{\hbox{\lower4pt\hbox{$\sim$}}}\hbox{$>$}}}}

\def\arcmin{\hbox{$^\prime$}}
\def\arcsec{\hbox{$^{\prime\prime}$}}

\def\farcm{\hbox{$.\mkern-4mu^\prime$}}

\renewcommand{\bibname}{References}

\begin{document}
\textheight=7.38in

\pagestyle{plain}
\begin{center}
{\bf\large Abstract}
\end{center}

\begin{center}
{\bf Cosmological Halos: A Search for the Ionized Intergalactic
Medium}
\end{center}

\medskip

\begin{center}
{\bf by}
\end{center}

\begin{center}
{\bf Robert M. Geller, Robert J. Sault, 
Robert Antonucci, Neil E. B. Killeen, Ron Ekers, and 
Ketan Desai}
\end{center}

\bigskip
\bigskip

Standard big bang nucleosynthesis predicts the
average baryon density of the Universe to
be a few percent of the critical density.
Only about one tenth of the predicted baryons have been
seen.
A plausible repository for the
missing baryons is in 
a diffuse ionized
intergalactic medium (IGM).
In an attempt to measure the IGM we
searched for Thomson-scattered
halos around strong high redshift radio sources.
Observations of the radio source 1935-692 were
made with the Australia Telescope Compact Array.
We assumed a uniform IGM, and isotropic
steady emission of 1935-692 for
a duration between $10^{7}-10^{8}$ years.
A model of the expected halo visibility function
was used in ${\chi}^{2}$ fits to place upper
limits on $\Omega_{\rm IGM}$.
The upper limits varied depending on the
methods used to characterize systematic
errors in the data.
The results
are 2$\sigma$ limits
of $\Omega_{\rm IGM}$ $<$ 0.65.
While not yet at the sensitivity
level to test primordial nucleosynthesis,
improvements on the technique will
probably allow this in future studies.

\thispagestyle{plain}
\tableofcontents
\thispagestyle{plain}

\chapter{Introduction}
\pagenumbering{arabic}

\textheight=7.81in
\pagestyle{fancyplain}
{\markright{Chapter 1:~~~Introduction}
\renewcommand{\chaptermark}[1]%
{\markboth{#1}{}}
\renewcommand{\sectionmark}[1]%
{\markright{Chapter 1:~~~Introduction}}
\lhead[\fancyplain{}\thepage]%
    {\fancyplain{}\rightmark}
\rhead[\fancyplain{}\leftmark]%
    {\fancyplain{}\thepage}
\cfoot{}


\def\gtwid{\mathrel{\raise.3ex\hbox{$>$\kern-.75em\lower1ex\hbox{$\sim$}}}}
\def\ltwid{\mathrel{\raise.3ex\hbox{$<$\kern-.75em\lower1ex\hbox{$\sim$}}}}
\def\H75{$H_o = 75$}
\def\Omgb{$\Omega_{b}$}
\def\OmgIGM{$\Omega_{\rm IGM}$}
\def\helium4{$^{4}$He}
\def\He3{$^{3}$He}
\def\Li7{$^{7}$Li}

\section{Standard Big Bang Nucleosynthesis}
 
According to the theory of the big bang the early universe was
extremely hot and dense.  About 1 second after the big bang the
temperature was around $10^{10}$ K and thermonuclear reactions
took place throughout the universe.
Within minutes, the light elements D, \He3, \helium4, and \Li7
were produced in quantities to remain unchanged until
the first stars formed.
The abundances of the primordial elements have been predicted
quantitatively and
these calculations are referred to as standard big bang
nucleosynthesis (SBBN)~\cite{sbbn91}.
Comparing the predictions of SBBN with observations
provides two great successes of the big
bang theory.
 
The main unknown quantity in SBBN is the baryon-to-photon
ratio.  Because we can
measure the CBR temperature very well, the main unknown becomes
the baryon density \Omgb~(expressed in units of the
critical density).  SBBN predicts that $\sim$ $24\%$ of the
mass of baryons after nucleosynthesis should be in the
form of \helium4~\cite{sbbn91}.
Because this result is roughly independent of \Omgb, it provides
the first direct quantitative prediction of the theory.
Measurements to determine the primordial abundance
of \helium4 are best made in metal-poor, highly ionized
extragalactic HII regions.  Low metallicity is preferred because
stellar processes can produce \helium4.  The results of several
observations
find a \helium4 abundance of $\sim~23\%$ (by mass fraction)~\cite{hel4}.
This agreement, within $4\%$ of theory and prediction for
\helium4, provides the strongest support for SBBN.
 
The abundances of D, \He3, and \Li7
can be calculated
as functions of \Omgb.
Although \Omgb~has
not been measured, there can only
be one universal value  
in a homogenous universe.
Consequently, for SBBN to be
correct, the observed (or inferred) primordial abundance
of each element must correspond to the same
value of \Omgb.
As we now discuss, the approximate agreement between the light
elements for a single range of {\Omgb}'s is the second greatest
success of SBBN.
 
Primordial deuterium is best measured in metal-poor
environments.  Stellar processes which destroy D should be
minimized in sites with low metallicity.
Several groups have measured the absorption of light from 
background quasars by intervening metal-poor
clouds containing deuterium.
The results are divided into high
abundances~\cite{song94,wamp96,webb97}
and low abundances~\cite{tytler96,burles96} 
separated by a factor of roughly 10.
Due to contamination of deuterium absorption
measurements by hydrogen, 
it is easier to get a false positive (high value) than
a false negative.
Recently Hogan (1997)~\cite{hogan97}
has estimated that hydrogen contamination could
reduce the high deuterium abundances by a
factor of two or more.
With this new correction, the high deuterium
measurements imply an \Omgb~$\sim$ 0.02 (for \H75).
The low deuterium measurements imply \Omgb~$\sim$ 0.05
(for \H75).

\He3 is produced in stars as they burn their primordial D
on the way to the main sequence.
There are other processes which both produce and
destroy \He3.  However, Yang et al. (1984)~\cite{yang84} showed
that the present sum of D+\He3 provides an upper
limit on their primordial abundance and thus places a
lower limit on \Omgb.  Walker et al. (1991)~\cite{sbbn91} used this method
to find a lower bound of \Omgb~$\gtwid$ 0.016 for
\H75.

Primordial \Li7 can be best measured in the atmospheres
of metal-poor, pop II halo stars.  Spite \& Spite (1984)~\cite{spite84} 
looked at the \Li7 abundance vs. surface temperature for
these stars.
They found a plateau in the \Li7 abundance 
for surface temperatures exceeding 5600 K.
Metal-poor stars are chosen because stellar processes
below the surface can deplete \Li7.  Low metallicity
implies low transport of material from inner regions
with depleted \Li7 abundances to the outer atmosphere.
Surface temperature is another indicator of potential
\Li7 depletion because high surface temperatures
result in shallower convection zones.
The plateau value of the \Li7 abundance is subject
to systematics such as the effective surface temperature
and depletion.  (The hot, metal-poor stars have minimal, but
non-zero depletion.)  Based on the measurements of
Spite \& Spite and Thorburn (1994)~\cite{thorburn94} , the
primordial \Li7 abundance,
by number, is $\sim$ 1.4x$10^{-10}$.
Copi, Schramm, \& Turner (1995)
(hereafter CST)~\cite{cst95} have assessed the
statistical and systematic uncertainties to obtain
a 2$\sigma$ range for the baryon density
inferred from \Li7.
The result is 0.006 $\ltwid$ \Omgb~$\ltwid$ 0.038 (for \H75). 

Combining the upper limits and possible
detections
(with reasonable error bars), bounds on \Omgb~can be
calculated for which all of the primordial abundances
are accounted for~\cite{cst95}.
CST compute the widest ranging 
concordance interval (for \H75) to
be 
0.010 $\ltwid$ \Omgb~$\ltwid$ 0.058.
The significance of this is that
SBBN can account for the primordial abundances
of D, \He3, and \Li7 for any
\Omgb~in the range above.
Stated another way, with only one free
parameter (\Omgb), SBBN can account
for all the light element abundances which span
about ten orders of magnitude.  
After predicting the correct \helium4 abundance, this
is the second greatest success of SBBN.

However, the study of big bang nucleosynthesis
does not end here. SBBN accounts for the primordial
light element abundances {\it only} if \Omgb~lies
in the concordance interval between 
1\% and 6\%.  In this sense, SBBN makes
a prediction for \Omgb, which is a rare quantitative
prediction on a fundamental cosmological parameter.

\section{Missing Baryons}

With SBBN's prediction in hand, we naturally want to
know the observed \Omgb~for comparison.  Persic \&
Salucci (1992)~\cite{persic92} estimated \Omgb~of the Universe
due to stars in galaxies and the hot gas in clusters.
They found \Omgb~$\sim$ 0.003, which is independent
of $H_o$.  CST predict for their lowest extreme that
\Omgb~$\gtwid$ 0.006.  This value is still twice the
baryon density observed by Persic \& Salucci.
For a more moderate comparison,
the central value predicted by CST is \Omgb~$\sim$ 0.034,
which leaves over 90\% of the baryons unaccounted for.
These missing baryons pose a serious dilema for SBBN.
One might decide that SBBN is `wrong'.  But, that would
be ignoring the success of the 24\% \helium4
abundance prediction and the consistency between the
other light element abundances.  
Consequently, the hypothesis has been made that all
of the baryons predicted by SBBN exist, but that
90\% have elluded detection.  These so-called `dark'
baryons have been a subject of great interest for over 30
years.

In addition to SBBN's assumption of
an initially homogenous mixture of protons and neutrons,
there are inhomogeneous nucleosynthesis
calculations as well.  These were originally motivated by
the hope that inhomogeneous nucleosynthesis could 
account for the primordial abundances with \Omgb~= 1.0.
Kurki-Suonio et al. (1990)~\cite{inhomo90} 
showed that inhomogeneous nucleosynthesis
is not consistent with \Omgb~= 1.0, but that consistency
is still possible for \Omgb~$<$~0.3.  Although SBBN may be
the simplest theory to account for the primordial abundances,
inhomogeneities may indicate a larger \Omgb~and even
more missing baryons.

One idea for baryonic dark matter is MACHOs (massive compact
halo objects). Stellar mass
MACHOs have been detected by microlensing
in the direction of the LMC~\cite{macho97}.
It has been proposed that these are white dwarfs
since they far exceed the census of ordinary stars.
In the opinion of Hogan (1997)~\cite{hogan97}, 
the parameters necessary to
extrapolate the observations into a global
density are too unconstrained.  Depending on the
values chosen for the galactic halo size and shape,
and the distance to the microlensing events,
the global baryon density due to MACHOs
could either be negligible, or dominate all
other forms.

\section{The Intergalactic Medium}

Another idea is that the missing baryons are
spread out in a somewhat uniform, mostly hydrogen
intergalactic medium (IGM).
In 1965 Gunn \& Peterson~\cite{gunn65} 
showed that any diffuse IGM
must be highly ionized.  
They reasoned that photons emitted shortward of Ly-$\alpha$ should
be scattered by diffuse neutral hydrogen as they
redshifted along a cosmologically distant line-of-sight.
Observing 3C9 at Z $\sim$ 2, they found the
neutral fraction of hydrogen in the IGM to 
be $\sim$ $10^{-6}$ (for \Omgb ~$\sim$ 0.05).
Even out to Z $\sim$ 4.5 the IGM remains highly
ionized~\cite{steidel87,webb92}.
Unfortunately, it is difficult to infer the total
baryon density from the 
small mass of inhomogeneously distributed neutrals
detected via Ly-$\alpha$ absorption.
With assumptions about the 
background ionizing flux and the
thermal history 
of the IGM, a range of 
0.006 $\ltwid$ \OmgIGM~$\ltwid$ 0.043
has been found (for \H75)~\cite{giall92}.

There have been several successful attempts to
measure an absorption trough for HeII~\cite{david96,
hog97,jakob95,rem98}.
Interpretation of the results is uncertain for a
variety of reasons.  It is not clear if the
HeII absortion is coming from Ly-$\alpha$
clouds or a diffuse IGM.
Even if that distinction could be made, assumptions
about the temperature and ionizing background
limit the ability to determine \Omgb.

\section{A Method to Measure \OmgIGM}

There is, perhaps, a method to determine \OmgIGM~without
the assumptions required when only the
neutral fraction is measured.
The dominant ionized component would result
in Thomson scattering of light which
should appear as ``fuzz'' around sources of
radiation in the cosmos.  In the case of
an isotropic point source of radiation in a uniform
ionized IGM, 
Thomson scattering would result in a
spherical halo centered on the source.
Clearly, the brightness of any measured
halo relative to the central source
is a measure of the density of the
scattering medium.  
By searching for halos, we hope to measure
the density of ionized gas on large scales
and thus directly measure \OmgIGM.

\afterpage\clearpage
\vfil\eject

\chapter{Cosmological Halos}

\pagestyle{fancyplain}
{\markright{Chapter 2:~~~Cosmological Halos}
\renewcommand{\chaptermark}[1]%
{\markboth{#1}{}}
\renewcommand{\sectionmark}[1]%
{\markright{Chapter 2:~~~Cosmological Halos}}
\lhead[\fancyplain{}\thepage]%
    {\fancyplain{}\rightmark}
\rhead[\fancyplain{}\leftmark]%
    {\fancyplain{}\thepage}
\cfoot{}


\def\gtwid{\mathrel{\raise.3ex\hbox{$>$\kern-.75em\lower1ex\hbox{$\sim$}}}}
\def\ltwid{\mathrel{\raise.3ex\hbox{$<$\kern-.75em\lower1ex\hbox{$\sim$}}}}

\section{The Sholomitskii Effect}
 
``Cosmological
Halos'' due to Thomson scattering
of radio waves in an ionized IGM
are expected around high Z sources.  Their properties have
been calculated in detail by 
Sholomitskii \& Yaskovich (1990)~\cite{sholo90}
and Sholomitskii (1991)~\cite{sholo91},
for uniform densities
and various assumptions about cosmology.  These
papers focus on sources with isotropic emission
assumed constant
over the source lifetime.
The halos are approximately 1/3 tangentially polarized
in surface brightness, although the total polarized
flux is zero if radial symmetry
is assumed.
The surface brightness distribution is $\sim 1/r$
and extends out to very large angular
distances limited in principle only by source lifetimes and the speed of
light.  For quasar emission ages between $10^7$~yr and $10^8$~yr,
halos at Z $\sim$ 3.5
could extend between $0.3^{\circ}$ $\ltwid$ $\Delta$$\theta$ $\ltwid
3.0^{\circ}$\footnotemark.
 
\footnotetext{Of course the constant-delay surface in 3-d is nearly
parabaloidal, and the
full-width at zero intensity
of the halo includes
the entire sky in principle.}
 
To estimate the total scattered halo flux, consider an expression
valid for small optical depth $\tau$,
 
{\centerline {
$ {\mathcal{S}}_{halo} \sim \tau {\mathcal{S}}_{o}$ ~ with ~
$\tau \sim n_{e} \sigma_{T} r_{c} $
}

Here ${\mathcal{S}}_{halo}$ is the scattered flux,
${\mathcal{S}}_{o}$ is the central source isotropic radio flux,
$n_{e}$ is the electron number density in the ionized IGM, $\sigma_{T}$
is the Thomson scattering cross section, and $r_{c}$ is the
light radius (emission age times the speed of light).
For $\Omega_{\rm IGM}=0.05$, $\Omega_o=1.0$, $H_o =75$,
Z = 3.5, and an assumed emission age of
$5$x$10^7$ yr, $\tau \sim 0.001$.  Therefore, a 1 Jy
high redshift radio source would
produce a $\sim$ 1 mJy halo spread out over
a size of order a degree.
Unfortunately, existing radio interferometers cannot
detect spatially smooth emission on scales of a degree
(at 20 cm) without
resolving out much of the flux.
For these large sizes
only 5-10$\%$ of the scattered flux is detected by typical
interferometers on the shortest baselines,
as can be seen in the next section.

\section{Model Halo Visibilities}

A two dimensional Fourier transform of an expected
brightness distribution results in a model halo
visibility function.
The appropriate brightness distributions are
derived by Sholomitskii (1992)~\cite{sholo92}.
(These are corrected
versions of Equations 8a and 8b in the 1991 paper\footnotemark.)
We performed numerical Fourier transforms
(using Mathematica)
to compute the visibility at many baselines between
100 and 700 $\lambda$'s (wavelengths).
These cover the relevant baselines for 20 cm
observations with antenna spacings between
31 m and 6 km.
To aid in signal-to-noise ratio (SNR) estimates and
fitting data in the $uv$ plane,
an analytic approximation was found
for the visibility function as follows:
 
\begin{equation}
{\mathcal{V}}^{S}(q,\phi) =
{\mathcal{S}}_{o} {(1 + Z)^{3} \over (1 + 3.5)^{3}}
{\left( {\Omega_{\rm IGM} \over 0.05 } \right)}
{\left( {{\omega}^{S}_{1} \over q }
+ {{\omega}^{S}_{2} \over q^{2} }
+ {{\omega}^{S}_{3} \over q^{3} } \right)}
\Theta^{S} (\phi) ~ mJy
\end{equation}
 
\footnotetext{There is an ambiguity in the corrected equations
which has been resolved through private communication with G.
Sholomitskii:  the argument of the $tan^{-1}$ function is
only the term $(1 - x^2)/(2x)$.}
 
Where ${\mathcal{V}}^{S}(q,\phi)$ is the visibility function
(in mJy)
for Stokes parameters S = I, Q, and U, as a function of baseline
$q=\sqrt{ u^{2} + v^{2}}$ (in wavelengths)
and $\phi $
is the baseline position angle
measured positive from $+u$ to $+v$
( ie, counter-clockwise as seen from the source; $\phi $ = 0
corresponds to the $+v$ axis, pointing North).
${\mathcal{S}}_{o}$ is the isotropic flux of the
central radio source in Jy.
A halo is not expected to produce Stokes V flux.
$\Theta^{S} (\phi)$ accounts for the azimuthal
dependence:
$\Theta^{I} (\phi)$ = 1 since the Stokes I visibility is
azimuthally symmetric;
$\Theta^{Q} (\phi) = -cos(2\phi) $, and
$\Theta^{U} (\phi) = -sin(2\phi) $.
The ${\omega}^{S}_{i}$ are three constants determined
from the fit for each Stokes parameter.
For ${\Omega}_{o}=1.0$ and $H_o =75$, the ${\omega}^{I}_{i}$ are
as follows:
${\omega}^{I}_{1}=13.6, ~ {\omega}^{I}_{2}=12.4$,
and ${\omega}^{I}_{3}=1.3$x$10^{4}$.
Similarly, ${\omega}^{Q}_{1}=4.5,~ {\omega}^{Q}_{2}=51.2$,
and ${\omega}^{Q}_{3}=-1.6$x$10^{4}$.
Because the radial dependencies of the Stokes Q and Stokes U
visibilities are equal, ${\omega}^{Q}_{i}$ = ${\omega}^{U}_{i}$.
Also, for a halo centered at the pointing center, the
visibility ${\mathcal{V}}^{S}(q,\phi)$ is purely real.
Additional numerical integration indicates that
a good model for ${\Omega}_{o}=0.2$ is obtained by
multiplying the above model by 1.55.
(This increase is due to the smaller angular size in an
open universe, which results in resolving out less
flux.)

It should be kept in mind that this simple fit
is valid only in the range $100 < q < 700$ $\lambda$'s.
To illustrate this point, recall that the
total polarized flux of an azimuthally symmetric
tangentially polarized signal is zero.
The total flux in a visibility function occurs
at $q=0$.  Therefore, the real visibility
function, for Stokes Q and U,
eventually rolls over to zero, whereas the approximation
above does not\footnotemark.
 
\footnotetext{Note: Someday when detection of
cosmological halos is {\it old hat},
there will be great interest in the
roll over
of the polarized visibility.
The baseline at which a polarization
visibility peaks and rolls over is a direct
measure of the halo size.
The size is simply a function of the emission age
and the speed of light.
Thus, measurement of the roll over can provide
emission ages of individual AGN's.
For emission ages between
$10^7$ yr and $10^8$ yr, the roll over occurs
between 25 $\lambda$'s and 50 $\lambda$'s, which is
somewhat accessible
for 90 cm observations.}
 
Figure 2.1 shows the visibility functions ${\mathcal{V}}^{I}(q,\phi)$
and ${\mathcal{V}}^{Q}(q,\phi$=$\pi/2)$ for
$\Omega_{\rm IGM}$ = 0.05,
${\mathcal{S}}_{o}$ = 1 Jy,
and Z = 3.5.  The Figure also assumes
$H_o =75$ and
${\Omega}_{o}=1.0$, but can be
multiplied by 1.55 to yield approximate
results for ${\Omega}_{o}=0.2$.
For 20 cm observations and a typical short
spacing of $\sim$ 30 m, q = 150 and
${\mathcal{V}}^{I}(150,\phi)$ $\sim$ 100 $\mu$Jy.


\begin{figure}
\begin{center}
\includegraphics[bb=27 170 564 695,width=5.4in]{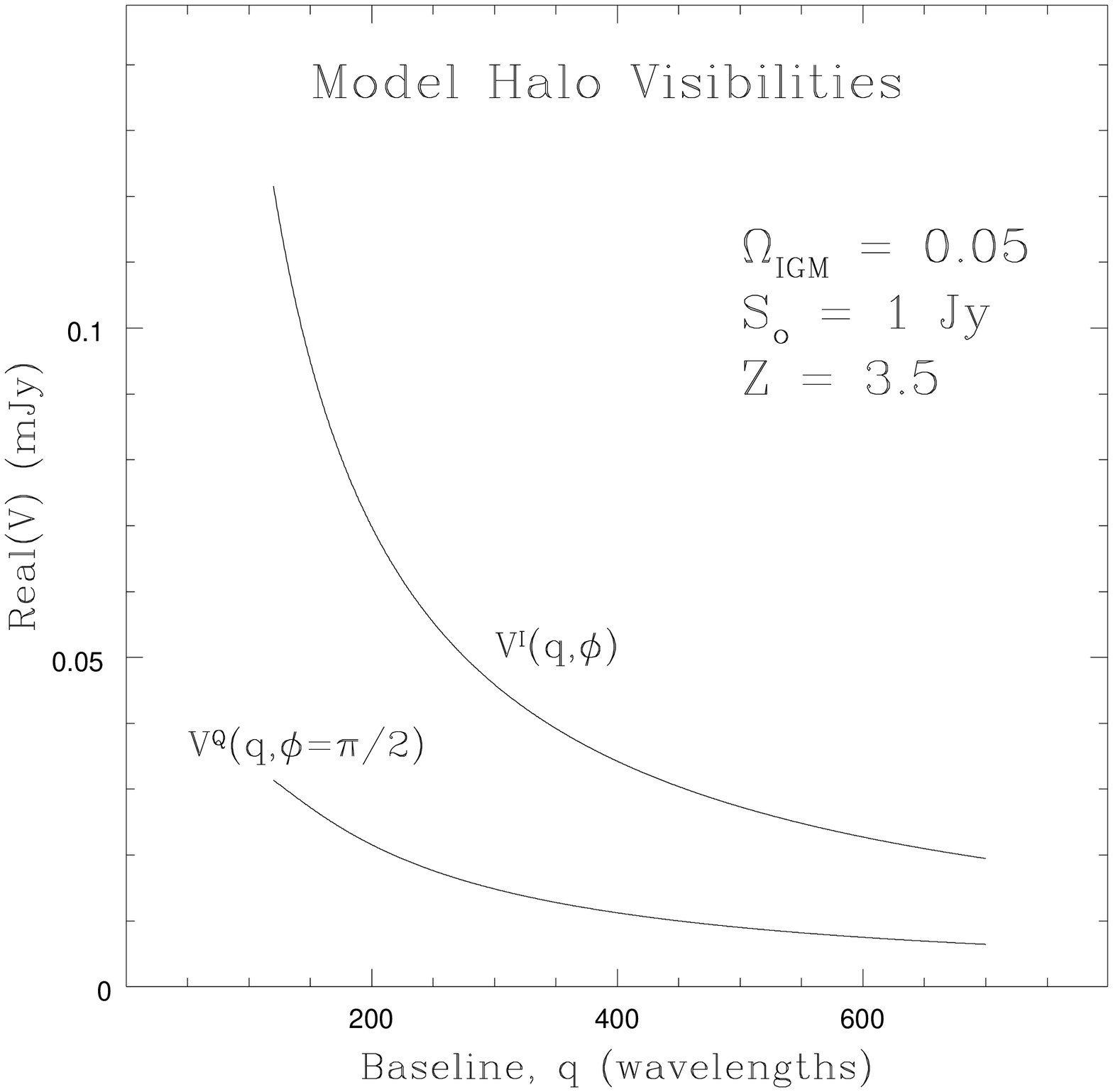}
\end{center}
\caption
{The model halo visibilities given by Equation 2.1.
The visibility functions ${\mathcal{V}}^{I}(q,\phi)$
and ${\mathcal{V}}^{Q}(q,\phi$=$\pi/2)$ for
$\Omega_{\rm IGM}$ = 0.05,
${\mathcal{S}}_{o}$ = 1 Jy,
and Z = 3.5 are shown.  The figure also assumes
$H_o =75$ and
${\Omega}_{o}=1.0$, but can be
multiplied by 1.55 to yield approximate
results for ${\Omega}_{o}=0.2$.}
\label{x1_model_1}
\end{figure}

\afterpage\clearpage
\vfil\eject

\chapter{Observing Strategy}

\pagestyle{fancyplain}
{\markright{Chapter 3:~~~Observing Strategy}
\renewcommand{\chaptermark}[1]%
{\markboth{#1}{}}
\renewcommand{\sectionmark}[1]%
{\markright{Chapter 3:~~~Observing Strategy}}
\lhead[\fancyplain{}\thepage]%
    {\fancyplain{}\rightmark}
\rhead[\fancyplain{}\leftmark]%
    {\fancyplain{}\thepage}
\cfoot{}


\def\gtwid{\mathrel{\raise.3ex\hbox{$>$\kern-.75em\lower1ex\hbox{$\sim$}}}}
\def\ltwid{\mathrel{\raise.3ex\hbox{$<$\kern-.75em\lower1ex\hbox{$\sim$}}}}

\section{Source Selection}
 
We wish to choose a source which maximizes the quantity
$$
{\mathcal{S}}_{halo} \propto {\mathcal{S}}_{o}*{(1+Z)}^3.
$$
where ${\mathcal{S}}_{o}$ is the 20 cm {\it isotropic} flux
from the central radio source, and the ${(1+Z)}^3$
redshift dependence accounts for the presumed epoch dependence
of the IGM density.
Unfortunately, observations of a given
source do not always allow you to infer the isotropic flux
with certainty.  Steep-spectrum radiation at centimeter
wavelengths from extended radio lobes is
mostly isotropic
on large scales, but these radio doubles are not the brightest
20 cm sources at high redshift.
To obtain a larger SNR, we must consider brighter sources.
 
The brightest sources which are plausibly nearly isotropic are
Gigahertz-Peaked Spectrum sources (GPS's)
(O'Dea 1997)~\cite{odea97}.
GPS sources are not very variable in flux density
and some have limits to their proper motion which are
sub luminal.
These two properties suggest that the sources are
not strongly Doppler boosted and that the
radio powers are intrinsic.
Our current halo search is around the GPS source 1935-692.
It has a 20 cm flux of 1.54 Jy, and a redshift of Z = 3.15.
(Consistent with O'Dea's interpretation of GPS's,
there is no evidence in VLBI images (J. Reynolds, PC)
of a beamed core-jet component for this object,
although VLBI data is sparse.
A map at 2290 MHz and 3 mas
resolution shows two partially resolved
blobs 8 mas apart.)

\section{Optimal Source Location}
 
A significant property of East-West arrays such as 
the ATCA and WSRT is that
equatorial objects undergo severe projection.  Projection
is a geometric foreshortening of the array baselines
which changes as the object
rises and sets.  In principle, projection is good for halo searches
starved for SNR because the halo visibility rises rapidly
at shorter baselines, but in practice {\it shadowing} may
limit any benefits.  Shadowing occurs when
the back of an antenna enters the field of view of
another antenna. Shadowing often occurs at low elevations
for low declination objects,
and affects the shortest baselines.  Although the shadowed
data can be flagged, this can mean throwing away
an unacceptably large portion of the data.
 
Another property of equatorial objects
observed with East-West arrays
is lower North-South image resolution;
all the baselines in the full synthesis are
very short in the North-South direction.
High ($\sim 5$ arcsec) resolution in all directions
is desirable
in the high resolution image used to characterize
confusing sources.
Although we originally favored sources at low declination
to benefit from an increased SNR
due to projection, we later
settled on objects at $|dec |$ $\gtwid$ $20^{\circ}$.

Finally, we attempted to stay at least $30^{\circ}$ away from
the Galactic plane.
Galactic synchrotron emission is stronger near
the plane and is likely to be associated with
Galactic Foreground Polarization (GFP),
to be discussed in section 3.8.
GFP can completely confuse halo observations
in polarization and should be carefully
considered in object selection.
For observations intended to look for the polarization
signature of halos, regions of bright Galactic
synchrotron emission should be avoided.  The Haslam 
survey~\cite{haslam82}
at 408 MHz 
should be consulted to evaluate the synchrotron background
of candidate sources.

 
1935-692 has a sufficiently high $|dec |$ to
preclude shadowing.  Its Galactic coordinates are
{\it l} $\sim$ 327, {\it b} $\sim -$29.  The average brightness temperature
in this region
at 408 MHz
is 30 K, and 1935-692 visually appears to be away
from brighter emission associated with the Galactic plane.

\section{NED Search}
 
With the above constraints in mind, we used 
NED\footnotemark~to produce a list
\footnotetext{The NASA/IPAC Extragalactic Database (NED) is
operated by the Jet Propulsion Laboratory,
California Institute of Technology, under contract with the
National Aeronautics and Space Administration.}
of radio sources with Z $\gtwid$ 2.  We also used NED to
find references for
spectral shape and other related information.
Based on the considerations outlined above,
1935-692 was the best source in the southern hemisphere.
1442+101 with ${\mathcal{S}}_{o}$ = 2.7 Jy
and Z = 3.5 is the best GPS source in the
northern hemisphere, but the low declination
makes it potentially better for the VLA than for an East-West array.


\section{Scattered Anisotropic Emission}
 
According to the beam model, the cores
of lobe-dominant sources may have
$\sim100$x the flux
of the isotropic lobes (at $\sim$ 4 cm rest wavelength) if seen
from the jet direction.  That is the case with typical observed
superluminal sources
(e.g. Antonucci \& Ulvestad 1985)~\cite{ski85}.
(This result might seem inconsistent
with the lack of observed $\sim 100$ Jy blazars at high redshift
as beamed versions of $\sim 1$ Jy doubles.
However, since such objects are selected from the very top
of the diffuse-radio-emission luminosity function, there
should be much less than one equivalent object in the
universe which beams towards Earth.)
We have estimated an
upper bound to the detectable scattered halo flux
assuming beamed emission in the sky plane
with directed jet flux $\sim100$x the isotropic flux.
The result is a signal roughly equal to the isotropic
component.
This approximate similarity between jet and halo
is due to the higher spatial frequency of the
jet resulting in less flux being resolved out
by the interferometer.
Note that this discussion is directly relevant to the large
double sources, rather than the GPS source observed here,
though similar considerations may apply.
 
\section{Possible Faraday Rotation in the IGM}
 
For polarization observations, Faraday rotation
can modify the initial tangentially
polarized signal.
To place upper limits on expected Faraday rotation
one must estimate a magnetic field strength and length
scale for field reversal.
(An electron number density corresponding to $\Omega_{\rm IGM}=0.05$
can be assumed.)
These quantities are needed for 30 Mpc scales
at Z $\sim$ 3.5,
but are unknowns.
Nonetheless, Vallee (1990)~\cite{vallee90}
has estimated $10^{-6}$G on
$\sim24$ Mpc scales and provides fiducial 
values for
estimating some of the necessary parameters.
 
\section{Possible Cluster Effects}
 
It is not known if 1935-692 is in a cluster.
If there is a cluster around our source, it
could produce an SZ effect, as well as
intracluster Thomson scattering and Faraday rotation.
As an illustrative calculation of the SZ effect,
we can take a Comptonization parameter
determined by Andreani et al. (1996)~\cite{SZ96}.
For a rich cluster at moderate redshift,
they found $y$ $\sim$ 1.3x$10^{-4}$.
The CMB in a ${10\arcmin}$x${10\arcmin}$
area 
(a generous value for a cluster solid angle)
is about 1.5 Jy.  Therefore, the negative
flux due to the SZ effect in a ${10\arcmin}$
cluster measured by an interferometer
could be $\sim$ (2$y$)(1.5 Jy) $\sim$ 400 $\mu$Jy.
In addition, this negative flux would be accompanied
by a positive flux due to intracluster Thomson scattering.
However,
as long as the cluster is $\ltwid 10\arcmin$, the
cluster emission can be spatially distinguished
from a halo on $20\arcmin$ (and larger) size scales.
For this reason, Faraday rotation
in the relatively dense intracluster gas
can also be distinguished from any large
size-scale halo polarizations.

\section{Sizes, Ages, and Time Dependence}
 
Emission ages of radio loud quasars are not known, but have been estimated
to be somewhere between $10^{7}$ and $10^{8}$ years, depending on the particular
source being modeled, its angular size, and
various poorly constrained modeling parameters
(Begelman, Blandford \& Rees 1984)~\cite{bbr84,scheuer95}.
The emission age determines the {\it light radius} of the halo and thus
its intrinsic size.
For an assumed age of $5$x$10^7$ years,
the light radius would be about 15 Mpc
which corresponds to a halo angular size of $\sim 1^{\circ}$
~(for Z = 3.5, $\Omega_o=1.0$, and $ H_o=75$).
It must be stated however that there is no guarantee
that a GPS source is old.  However, the age would need to be
$<<$ $10^{7}$~yr to reduce the halo fluxes
substantially on the relevant baselines.
O'dea (1997) reviews the arguments for and against
young and old GPS's.
 
Interferometers cannot
measure flux 
which is smooth on scales greater than a characteristic scale set
by the observing wavelength and the shortest baseline.
For 20 cm observations and a typical shortest spacing of
30 meters, the maximum size scale detectable is about 20 arcmin.
This means much of the flux within a $1^{\circ}$ halo is ``resolved out''
and therefore undetected.  Thus,
on available baselines, the halo observations
and models are relatively insensitive to assumed ages
between values of $10^{7}$ and $10^{8}$ years.
Therefore, our assumed age
of $5$x$10^7$ years is fairly robust
for the range of available baselines.
 
Following Sholomitskii, we assume a constant
radio luminosity for 1935-692 for the duration of its
emission age.
Given that our GPS source is at an unknown evolutionary
state, the time dependence of the luminosity
is unknown, although flux selection would result
in catching sources near peak output.
This would tend to cause a sharper decline
in the surface brightness with radius
than the $\sim 1/r$ suggested above.
 
Finally, on time scales $10^{11}$ times shorter, variability can
cause other problems for radio observations.
A central component
can vary
over the course of the observations (3 months),
violating a fundamental assumption of interferometry.
The requirement of time independence derives
from the deconvolution, which uses a beam pattern from a
Fourier
transform of the array positions during the {\it entire} observation.
The only check that can be made is to image the data in
short time segments,
subject to the limitation that good {\it uv} coverage
must be present.
The problem is potentially serious because data for sparse
arrays is collected over timescales of weeks to months in order
to get better {\it uv} coverage from several array configurations.
Inevitably, some of the confusing sources will vary, but attention
should be paid
to avoiding
selecting central sources for which previous observations
suggest variability.
(Even aside from interferometric considerations, variability
could be a dangerous sign of anisotropic flux.)
For 1935-692, we found no evidence of time dependence
in the data taken over a span of three months.
 
\section{Confusion}
 
Confusion generally refers to any unwanted emission in the field of view.
This emission usually comes in the form of numerous radio galaxies
and quasars, a few
of which can be quite strong.  These sources can be
imaged at high resolution and subtracted from the short spacing data,
but image artifacts often limit the accuracy of the subtraction.
At 20 cm the average flux of the brightest confusing source
per primary beam is typically 100 mJy.
It is best to avoid fields with peak confusion beyond
several hundred mJy, and beneficial if the field has a lower
than average
peak confusing flux.
 
In addition to confusion in total intensity, there may be
a potentially significant polarized contribution.
On average, each (integrated)
source will be only
a few percent polarized,
reducing the confusion for
halo searches in polarized flux.  Unfortunately, there is
another source of confusion which can
dominate any faint halo
polarization signal:  Galactic Foreground
Polarization (GFP)~\cite{wiering93}.
Observationally, GFP appears as
lumpy structures in the Stokes Q and U maps varying in size
from at least 10 arcmin
to several degrees.
There appears to be no detected Stokes I
emission associated with GFP.
A simple model developed to account for these
characteristics is a relatively uniform
background of polarized Galactic synchrotron emission
with an intervening Faraday screen.  The background
is too smooth for detection in Stokes I, and the Stokes Q and
U signals are broken up into higher (observable) spatial
frequencies by the Faraday screen.
Although GFP appears to be concentrated in regions of
strong Galactic synchrotron emission (and thus, our
recommendation regarding source position in section 3.2),
GFP can still occur away from the Galactic plane.
A short integration, low
resolution polarization image should be made of any field
to check against GFP before any long observations are made.
Unlike typical small-scale confusion, GFP cannot be
modeled and subtracted from the shortest baselines
without introducing errors on the same size scale as, and
indistinguishable from, a halo signal.
Since some telecopes,
such as the VLA, require a sacrifice in sensitivity to
correlate polarization information
(in spectral line mode), it should be determined
beforehand if a field has GFP which might
confuse a polarized halo search.
As discussed in section 4.5, we found polarized
emission in the field
of 1935-692 consistent with GFP.

\afterpage\clearpage
\vfil\eject

\chapter{ATCA Observations}

\pagestyle{fancyplain}
{\markright{Chapter 4:~~~ATCA Observations}
\renewcommand{\chaptermark}[1]%
{\markboth{#1}{}}
\renewcommand{\sectionmark}[1]%
{\markright{Chapter 4:~~~ATCA Observations}}
\lhead[\fancyplain{}\thepage]%
    {\fancyplain{}\rightmark}
\rhead[\fancyplain{}\leftmark]%
    {\fancyplain{}\thepage}
\cfoot{}


\def\gtwid{\mathrel{\raise.3ex\hbox{$>$\kern-.75em\lower1ex\hbox{$\sim$}}}}
\def\ltwid{\mathrel{\raise.3ex\hbox{$<$\kern-.75em\lower1ex\hbox{$\sim$}}}}

\section{Detection Strategy at the ATCA}
 
Detection of a large size-scale, faint signal in the presence of
foreground (confusing) sources requires a low-noise instrument with both
high and low resolution, and observations at wavelengths
with low background contributions.
Because the proposed faint halo surrounds a bright central radio source,
a high dynamic range is also required.
The instrument which
best satisfies these criteria is a
compact radio telescope interferometric
array.  For our observations, we used the Australia Telescope
Compact Array (ATCA) which has 6 antennas movable over a 6 km
rail track.  The longer baselines obtain high resolution
images of the numerous foreground radio sources, and the shortest
baselines around 31 m are sensitive to 20 arcmin structure
(appropriate for halos) at
our observing wavelength of 20 cm.
The ATCA has two features which are advantageous
for halo searches.  First, the availability of
four 31 m spacings increases the
20 arcmin scale sensitivity, and allows redundancy
to be used in gain
calibrations.
Second, the ATCA can correlate all four
Stokes polarization parameters over a 128 MHz bandwidth.
Because a halo would be approximately 1/3 polarized,
it is desirable to correlate polarization information.

 
The ATCA's short spacings of 31 m, 61 m, 92 m, and 122 m are available
in a configuration called the ``122 meter'' array.
These spacings are sensitive to
large scale structure up to about 20 arcmin.
They can sample a faint halo, but they also sample all the smaller
confusing sources.  To remove the effects of the confusing sources,
high resolution images are made from several larger array
configurations with baselines reaching up to 6 km (6 arcsec
synthesized beam).
Models of the smaller confusing sources can be made using
CLEAN algorithms and subtracted from the low resolution 122 meter
data.  Ideally, all the confusing sources are small enough to be
sampled and properly modeled by the high resolution arrays
(up to about 5 arcmin).
Subtracting the model made with high resolution data from
the low resolution 122 meter array data should then result in
low resolution data containing only the response to
the proposed faint halo.  There are two main limitations
to this procedure.
First, if there is confusing emission which is larger than
about 5 arcmin, it can not easily be modeled and subtracted
from the 122 meter array ``halo data''.  Second, in addition to
thermal noise, there are systematic errors
in characterizing and subtracting the confusing sources
which can
be larger then the halo signal.  

\section{The Observations}

The observations were made during mostly
night time hours to minimize phase errors.
The two available 128 MHz observing bands were centered on
1344 and 1432 MHz, which is relatively free of
terrestrial interference. The ATCA correlator subdivides each of the
128 MHz bands into 33 channels. The 33 channels in each band have
effective bandwidths of 8~MHz, and 
adjacent channels overlap by 50\%. 
Following standard ATCA procedure,
we discarded half of the
overlaping channels without any lost is sensitivity.
The observations provided full polarization
information.

For good $uv$~coverage in the high resolution configurations
we observed with four different array configurations
over a total of 9 nights (see Table 4.1).
The array configurations
gave a good sampling of baselines from 77~m to 6~km.
Table 4.1 also lists the 10 nights taken in the low
resolution array configuration (the so-called 122 m array). Here five of the
six antennas are placed at increments of 31~m, with the sixth antenna being
approximately 6~km away.

\begin{table}
\caption{ATCA Observations of 1935-692}
\begin{threeparttable}
\begin{tabular}[hptb]{cccc}
\hline
\hline
Array & Resolution\tnote{i} & Dates &
Observation time (hours)\tnote{j}\\
\hline
6B & high & 08/31/96 to 09/01/96 & 14 + 15
 \\
6C & high &08/09-10/96 & 12 + 15
 \\
1.5A &high& 10/25-27/96 & 12 + 10 + 13
 \\
0.75A &high& 11/08-09/96 & 11 + 13
 \\
122B &low& 08/20-29/96 & 14/night\\
\hline
\label{tbl-1}
\end{tabular}
\begin{tablenotes}
\small{
\item[i] High resolution arrays have a FWHM beam between 6 and 10
arcsec, and the low resolution array has 
a FWHM beam about 5 arcmin.
\item[j] Observations made during mostly night time hours.
\\
}
\end{tablenotes}
\end{threeparttable}
\end{table}

A potential problem with interferometers is
that ``DC offsets'' in the signal path can give rise to an errant,
possibly time dependent,
signal which mimics a source at the delay center. DC offsets can
arise in several parts of the system. Many radio interferometers
reduce their effect by phase switching (i.e. modulating the
astronomical signal by, say, a Walsh function at an early stage in the
system path, and demodulating it at a late stage), with the modulation
period typically being a small fraction of a second.  Although the
ATCA's design (digitizing quite early in the signal path) eliminates
many potential components where DC offsets could be produced, the
potential for DC offsets is not completely eliminated. 
Thus, ATCA observations are
occasionally affected by weak DC offsets. Given the  high dynamic range
required for this experiment, we have used two techniques to minimize
the potential for DC offsets affecting the data. Firstly we have
implemented a ``poor man's phase switching'' with software in the on-line
system. We modulate the instrumental phase by Walsh functions with a period
of 10~s (the system integration time), and remove this phase in the off-line
software. An alternative second approach is to move the delay
center of the observations away from 1935-692, 
so that any errant signal is well
away
from the region of interest.

For the high resolution observations, we
moved the delay center
$30''$ south of the pointing center (the pointing center remained on
1935-692). $30''$ was chosen because at several synthesized
beams away, an errant source in 
the image can be distinguished from 1935-692.
Data reduction later revealed that this offset was
unnecessary, but it is an intermittent
danger worth avoiding. The poor man's phase switching was not used for
these observations.
 
In the low resolution observations,
the potential problem of
DC offsets at the delay center is harder to address.
Again, one choice is to move the delay center
away from the pointing center.
However, to move the delay center
sufficiently far from the main lobe
of the 31~m spacing requires an offset of order a degree
which would result in some bandwidth and time smearing of the data.
(The data to the sixth antenna would be very badly affected.)
To hedge our bets, we
observed 5 nights using phase switching,
and 5 nights with a $\sim$ 1 degree southern offset in the delay
center.

A bright flux and phase calibrator,
1934-638, was interleaved
each night for 5 minutes of every
half hour.
1934-638, like 1935-692, is a GPS source, with
$ {\mathcal{S}}_{20cm}$ = 15 Jy and Z = 0.2.
Although self-calibration is possible
from 1935-692, we wanted a calibrator observed
as similarly as possible to the target source.
We found the calibrator observation to be
invaluable for investigating systematic errors
and highly recommend its inclusion in any
observing program.
We chose a duty cycle which results
in a ratio of peak flux to thermal
rms comparable to the target source.
In the extreme case,
if we saw a halo around 1935-692, we would want to check
that we did {\it not} see one around the calibrator.
This is because
Equation 2.1 predicts no detectable halo at the low
redshift of the calibrator.
This test is meaningful only if the
calibrator has the same $uv$ coverage,
phase offsets or phase switching, etc., as the target source.

\section{Data Reduction}
 
Data reduction was performed in the Miriad system 
(see Sault et al. 1995~\cite{sault95}).
Initial estimates of
feed-based gains, bandpass functions
and polarization leakages were determined from the observations of
1934-638. The bandpass
functions and polarization leakages were assumed constant during the
course of a night's observation, whereas the complex gains were 
calculated for
every 30~min.
Sault et al. (1991)~\cite{sault91}
describe the gain and polarization calibration
technique for the ATCA.

Discarding the edge channels and every second channel, we are left with 13
channels 
8~MHz wide in
each of the two bands. As the fractional width of each band ($\sim 8$\%) is
appreciable and because we wish to combine the two bands in the
imaging, we needed to use a ``multi-frequency synthesis'' approach to the
imaging and deconvolution.
Basically multi-frequency synthesis is the practice of forming an
image from data measured at a variety of frequencies (i.e. in our
case, from two bands with 13 channels per band). Multi-frequency
imaging differs little from conventional interferometric imaging.
We used the so-called grid-and-FFT method, which is nearly universally
used\footnotemark. 
\footnotetext{We did not follow the AIPS procedure of using 
direct Fourier tranforms (DFT's)
for the first few CLEAN cycles.  When software and hardware permit,
it would be ideal to combine DFT's with multi-frequency synthesis.}
In multi-frequency imaging, the visibility data is gridded at
its true $uv$ coordinate rather than the $uv$ coordinate corresponding
to some average frequency (note that $uv$ coordinate is measured in
wavelengths, and so it is a function of both the spacing and the observing
frequency). This has the advantages of reducing bandwidth smearing effects
and improving the $uv$ coverage.
 
Multi-frequency synthesis, however, does have its drawbacks.
Because we have used multiple frequencies in the imaging, an error is
introduced into the images because of the varying source flux density
with frequency. Conway et al. (1990)~\cite{con90} 
has analyzed issues involved in
using multi-frequency synthesis. In particular, they show that it is
possible to correct for the effects of non-zero spectral index when
deconvolving images. Sault \& Wieringa (1994)~\cite{sault94}
extend this work, and
describe the deconvolution algorithm that we have used.
The deconvolution algorithm is an extension of the CLEAN
algorithm~\cite{hog74}, 
whereby the image is decomposed into a
collection of point sources. Unlike the classical CLEAN algorithm, the
multi-frequency variants also estimate the spectral index of the
source and correct for the error caused by non-zero spectral index.
The technique of spectral index estimation and correction assumes
that the spectrum of a source is linear (or at least well approximated
as linear) over the spread of frequencies. Following the suggestion of
Conway et al. (1990) and Sault \& Wieringa (1994), our calibration of the
visibility data has forced the spectrum of 1935-695 to be linear, and thus
the linear assumption in the deconvolution step is a good one. 
Generally, a remaining
spectral index error results from differences between the spectral index
of the confusing sources and 1935-692. The residual spectral errors
will be apparent around the {\em second brightest} source in our images
(as seen around source ``a'' in Fig.~4.1). 

Regular observations of a calibrator (1934-638) are quite inadequate
to achieve the dynamic ranges required by this experiment. We have used
the techniques of self and redundant calibration to 
improve 
substantially
the calibration and the ultimate dynamic range. To understand
these techniques, we note that the main calibration
errors are antenna-based (strictly we should call them feed-based) gains, and
that there are $N$ of these ($N$ being the number of antennas).
However an interferometer array produces $N(N-1)/2$ complex measurements per
integration. For large $N$, the number of gains is modest compared with
the number of measurements.
Particularly,
if we add in some extra constraints on the data, it is possible to
deduce
simultaneously the antenna gains and the corrected visibilities.
 
In the case of self-calibration, the extra constraints are effectively
image positivity and that the sky is relatively sparse 
with bounded support. These constraints are implemented indirectly
by imaging and deconvolving the data to produce a model of the sky.
Model visibilities are then produced from this model of the sky, and
a least-squares solution is found for the antenna gains which minimizes
the difference between the model and calibrated visibilities.
Pearson and Readhead (1984)~\cite{pear84} 
give a thorough review of the self-calibration
technique.
 
Redundant calibration
(see Noordam \& de Bruyn 1982~\cite{noord82}) 
is applicable
where the interferometer array
measures the same spacing using different pairs of antennas (i.e. there
are redundant spacings). In this case, the extra constraint is that
the calibrated visibilities on the redundant spacings must agree.
 
For the high resolution data, there are no simultaneous redundant baselines,
and a conventional self-calibration procedure was used to obtain the best
estimates of antenna gains. Eight iterations of
Imaging -$>$ Cleaning -$>$ Self-Calibration were performed
(5 with phase only, and 3 with phase and amplitude).
Because we have
sampled the $uv$ plane very well with a good collection of array configurations,
a high quality model can be produced. In the self-calibration process, we
assumed the gains were independent of channel and frequency, and so all the
channel data at a given instant could be used in determining the gain
solution.
In producing the model visibilities at the different channels,
we have included the effect of the spectral index deduced from the
multi-frequency processing as well, of course, as the changing $uv$ coordinate
of the different channels.
 
The low-resolution data are quite a different matter. We could simply use
the model derived from the high-resolution data for the model for
the low-resolution self-calibration. A characteristic of self-calibration is
that it will tend to force the data to look like the model (whether the
model is correct or not). As the high resolution data
has essentially no sensitivity to a potential halo, simply using the
high resolution model risks suppressing the halo. Additionally the 122 m
array is a poor imaging array, and so self-calibration's image-based
constraints are less useful here. However the 122 m array is highly
redundant (there are 
four 31~m spacings, three 61~m spacings and two 92~m spacings).
We
have used a hybrid redundant/self-calibration scheme for these data. In
an algorithm similar to that of Wieringa (1992)~\cite{wier1992}, 
for each integration we
found antenna gains which minimized
\begin{equation}
\epsilon^2 =
(1-\lambda)\sum_{i,j}|M_{ij} - g_i^{\hbox{}} g_j^\ast V_{ij}|^2 +
        \lambda\sum_{i,j,k,l}
        |g_i^{\hbox{}} g_j^\ast V_{ij} - g_k^{\hbox{}} g_l^\ast V_{kl}|^2.
\end{equation}
The two summations are self and redundant calibration terms respectively,
with the first summation being over all baselines, and the second being over
all redundant pairs of baselines.
$M_{ij}$ and $V_{ij}$ are the model sky and measured visibility
for baseline $i,j$ and the $g_i$ are the antenna gains that are being
determined. $\lambda$ is a constant which adjusts the 
relative weight given to the
self and redundant calibration. We typically used a value of 0.9
to give the redundant calibration more of the weight.

The results 
of data reductions with self-calibration, as opposed to
redundancy calibration, can be compared.
The next chapter discusses the results of data reduction in detail, 
but we can simply refer here to figures in Chapter 5
which contain the results of various
calibration techniques.
In Figure 5.2, the top plot shows three groups of points.
There are four points on the left at a baseline
of $\sim$ 140 wavelengths.  These are from four redundant
baselines produced by antennas separated by about 31 meters.
The other two groups of points are produced by
two additional redundant spacings.  Because equal
baselines should sample the exact same brightness
distribution, they should only differ by random thermal
errors.
The thermal error bars are plotted for each point
and it can be seen that the scatter among each set of baselines is 
consistent with thermal errors.  All of the data in 
Figure 5.2 has had a redundancy calibration
performed after initial applications of 
self-calibration.
For comparison, Figure 5.3 a) shows the same data
as the top plot in Figure 5.2, but with only
self-calibration performed.
It can be seen in Figure 5.3 a) that the
scatter among the baselines is not consistent
with random thermal errors.  Most of the
points are 
at least several standard deviations from their means.
Although not every night plotted in Figure 5.2 is
as consistent with thermal errors as the first
night, redundancy calibration showed a
significant improvement in the internal
consistency of the data.  For this reason,
we performed redundancy calibrations
on all 
low resolution data processed for halo analysis.

One of the limiting errors in our data is apparent in
Figure~4.1 where 1935-692 is surrounded by fine-scale rings.
Some not insignificant amount of investigation showed that these rings are
caused by phase noise in the local oscillator system.
This phase noise leads to a small amount of amplitude decorrelation.
The antenna-based phase noise is more correlated between 
some antennas 
than others.
This results in the amount
of decorrelation being baseline based (not antenna based), i.e. it is a
non-closing error. As the resultant errors are believed to be equivalent to
time-independent, real-valued, baseline-based gain factors,
we attempted to deduce these factors
from the observations of 1934-638. 
However applying these factors to the 1935-692
data failed to reduce the errors. 
We think this is because our antenna-based
errors are larger than our 
baseline-based errors.  


\section{High Resolution Imaging Results (I,Q,U)}
 
There are two purposes of the high resolution image:
calibration of the low resolution data and
subtraction of confusing sources from the low
resolution data.
The high resolution image is used to 
self and redundancy calibrate
the 122 meter
array data.
Then, to search for a faint halo,
CLEAN models of
confusing sources made from the high resolution image
are subtracted from the
low resolution data.
For both purposes an accurate model is required and
this in turn demands a high dynamic range (DR) image.
Consider Figure 2.1 for the model halo visibilities
around a 1 Jy central quasar at Z = 3.5.  The 31 m spacing
has $\sim$ 105 $\mu$Jy of flux.
It is difficult to estimate the required DR to detect an extended halo
because it depends on the spatial
distribution of resulting artifacts.  Nonetheless,
for each source in the field to be modeled to better than
105 $\mu$Jy, a minimum DR of $\sim$ 10,000:1
would be needed.
 
Scattering models also predict an approximately $1/3$
tangentially polarized halo.
Therefore, we made high resolution images of
both Stokes Q and U for the field of 1935-692.
As expected, the confusion from radio galaxies
and quasars is much lower in polarization
compared to total intensity.  Because 1935-692
is only $\sim 1\%$ polarized, there is no
significant DR requirement in polarization images.

Figure 4.1 is the Stokes I image made from combining
all 9 nights of data in the four high resolution
configurations listed in Table 4.1.
1935-692 is the brightest source with 1.54 Jy and
appears at the center of the image.
Table 4.2
lists the Stokes I, Q, and U for the next 5
brightest confusing sources labeled a, b, c, d, and e in the image.
Computing the noise in a large box within
the primary beam which excludes
the brighter artifacts, our image has
an rms of $\sim$ 20 $\mu$Jy.  Therefore we achieved
an average DR of 77,000:1 for this image.
However, artifacts in the image are often $\sim$ 100 $\mu$Jy,
reducing the effective DR in those regions.
The noise in the Stokes Q and U images
is $\sim$ 15 $\mu$Jy.  DR artifacts in Q and U
are negligible.
The relevance  of high DR to halo searches
is best understood in the context of
systematic errors manifest in the $uv$ plane.
Therefore, we revisit the subject of high DR in
chapter 5 where these systematic errors are
discussed.
 
\begin{figure}[p!]
\begin{center}
\end{center}
\caption
{(This figure has been removed to reduce size.
It can be obtained by contacting Robert Antonucci
at ski@ginger.physics.ucsb.edu):
The Stokes I image made from combining
all 9 nights of data in the four high resolution
configurations listed in Table 4.1.
1935-692 is the brightest source at 1.54 Jy, and
appears at the center of the image.
Table 4.2
lists the Stokes I, Q, and U, for the next 5
brightest confusing sources labeled a, b, c, d, and e in the image.}
\label{x2_1935all_imfrst4}
\end{figure}

\begin{table}
\caption{Sources in 1935-692 High Resolution Image}
\begin{threeparttable}
\begin{tabular}[hptb]{cccc}
\hline
\hline
Source & I Flux (mJy) &  Q Flux\tnote{i} & U Flux\tnote{j}\\
\hline
1935-692\tnote{j} & 1540 & 8.5 & 15.2\\
a & 192 & -0.112$<Q<$0.100 & -0.063$<U<$0.060\\
b & 36 & -1.03 & 0.309\\
c & 21 & -0.035$<Q<$0.041 & -0.056$<U<$0.046\\
d & 14 & -0.367 & -0.047$<U<$0.052\\
e & 10 & -0.090 & -0.044$<U<$0.044\\
\hline
\label{tbl-2}
\end{tabular}
\begin{tablenotes}
\small{
\item[i] Bounded fluxes are reported where no reliable
detection was made.  The varying bounds reflect the artifacts around
each source location. 
\item[j] For this source only, a reliable V flux
of -2 mJy was measured.
\\
}
\end{tablenotes}
\end{threeparttable}
\end{table}

The grey scale for Figure 4.1 was chosen to bring out the ring
artifacts around several of the sources.
The rings are due to errors in the synthesized beam pattern
deconvolution.
Object variability, as well as time dependent systematic errors
can lead to artifacts.
They can also result from
misuse of, or limitations in, the data reduction software.
Multi-frequency synthesis
is less effective for sources near the edge of the
primary beam.  An example of this is source ``b'' in
Figure 4.1 where an error in the spectral index produces
visible rings of $\sim$ 100 $\mu$Jy in amplitude.

\section{Low Resolution Imaging Results (I,Q,U)}
 
Discussion of the Stokes I data requires
an analysis of the systematic errors.
We defer this to the next section
and focus on the Stokes Q and U results here.
We subtracted high resolution CLEAN models
of the polarized sources from the low resolution
Stokes Q and U data to search for a halo.
We found large scale polarized emission
in both Stokes Q and U, but it appears to
be GFP (see section 3.8) rather than
halo emission.
Figures 4.2 and 4.3 are
low resolution maps of Stokes Q and U
respectively.  
The high resolution CLEAN models of
the sources were subtracted before imaging,
but the images themselves are uncleaned.
The synthesized beam of the low resolution array
is 5$\farcm$5 x 3$\farcm$9
(with a PA of $-0.2$ degrees).
The Stokes Q flux is $-15$ mJy
in a 20$\arcmin$ x 10$\arcmin$ box oriented East-West and
centered on the peak emission at (800$\arcsec,-250\arcsec$).
The Stokes U flux is $-10$ mJy
in the same size box when centered on the peak emission at
(500$\arcsec,-200\arcsec$).
Clearly, the polarized emission is not centered around 1935-692.
The flux is unlikely to be due to a halo since
the Stokes Q and U are each greater than
the Stokes I in the same 20$\arcmin$ x 10$\arcmin$ box.
Halo models unequivocally predict that detected Stokes I
should be greater than any scattered polarized flux.
On the other hand, GFP is usually characterized
by having polarization
without total intensity.
None of the Stokes Q, U, or I, high resolution images
have any apparent source at the location
of the peak emission.
To check against systematic errors, we carefully
checked the field of 1934-638 for similar
features.  We found no apparent emission
consistent GFP or strong systematics.
The large scale polarized confusion is
about 100 times our expected halo flux for
an $\Omega_{\rm IGM}=0.05$.  As a result, we were unable to
search for halos or place meaningful limits
on $\Omega_{\rm IGM}$ using polarized flux.

\begin{figure}
\begin{center}
\includegraphics[bb=35 112 567 711,width=4.9in,angle=-90]{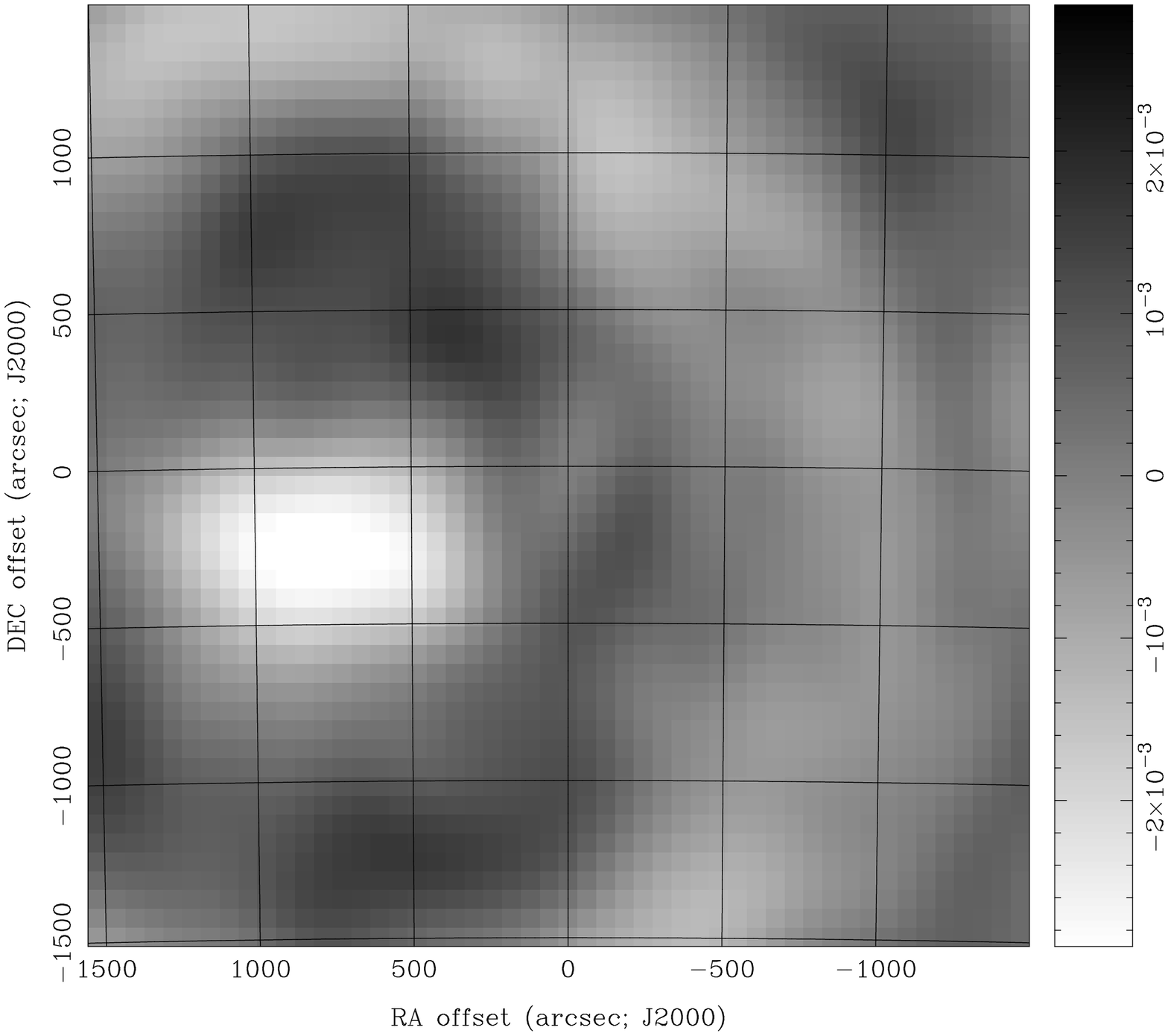}
\end{center}
\caption
{The Stokes Q
low resolution map of 1935-692.
The image is uncleaned
and the high resolution CLEAN models of
the sources were subtracted before imaging.
The synthesized beam of the low resolution array
is 5$\farcm$5 x 3$\farcm$9 (with a PA of $-0.2$ degrees).
In a 20$\arcmin$ x 10$\arcmin$ box
centered on the peak emission 
at (800$\arcsec,-250\arcsec$),
the flux is about
$-15$ mJy.} 
\label{x3a_1935_subHRq_qmp1_1}
\end{figure}

\begin{figure}
\begin{center}
\includegraphics[bb=35 112 567 711,width=4.9in,angle=-90]{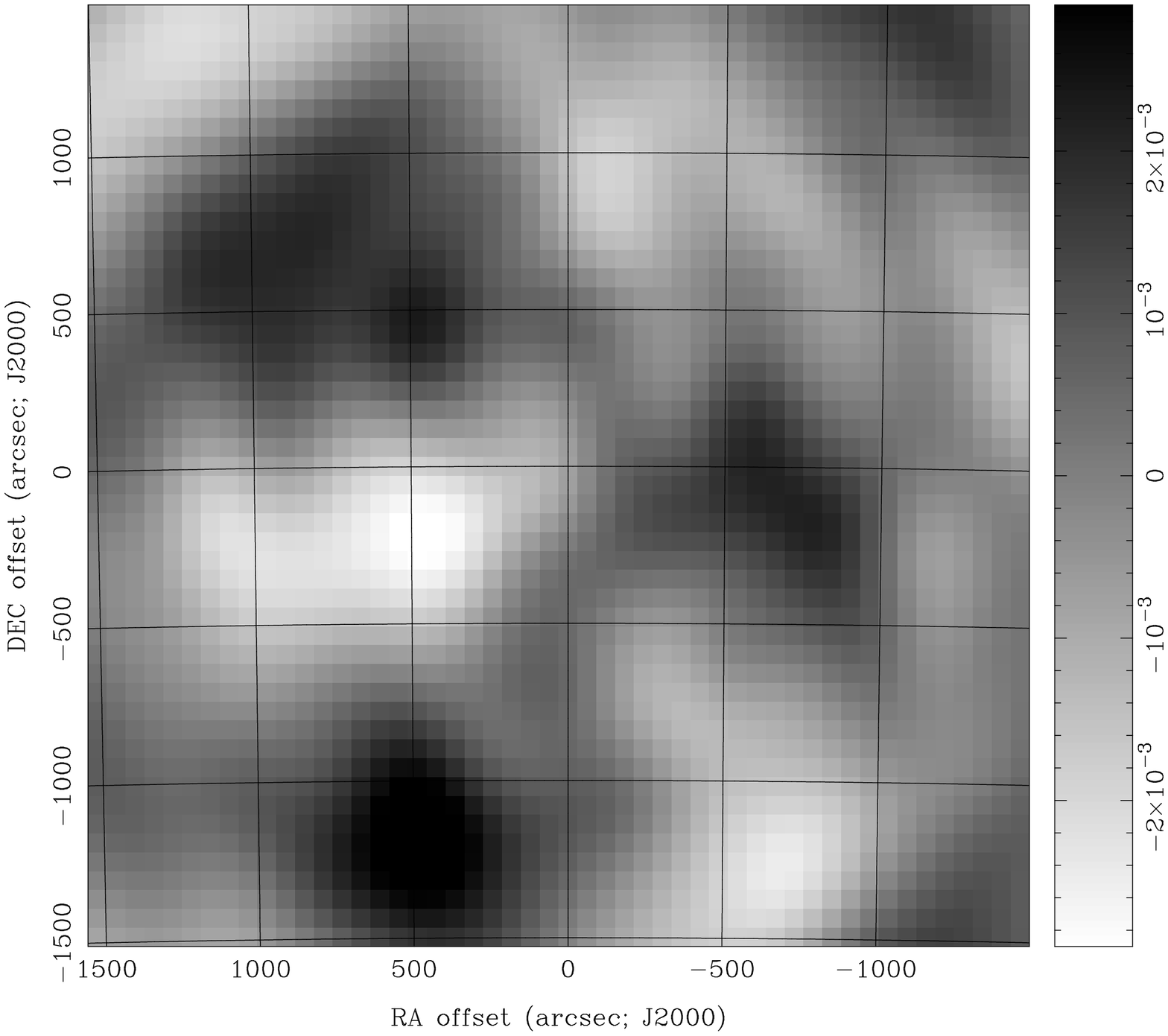}
\end{center}
\caption
{The Stokes U
low resolution map of 1935-692.
The image is uncleaned
and the high resolution CLEAN models of
the sources were subtracted before imaging.
The synthesized beam of the low resolution array
is 5$\farcm$5 x 3$\farcm$9 (with a PA of $-0.2$ degrees).
In a 20$\arcmin$ x 10$\arcmin$ box
centered on the peak emission 
at (500$\arcsec,-200\arcsec$),
the flux is about
$-10$ mJy.} 
\label{x3b_1935_subHRu_ump1_1}
\end{figure}

It is worth mentioning that we may
have detected
a low resolution Stokes I counterpart
to the polarization signal
in the field of 1935-692.
We measure about 4 mJy of Stokes I flux at the location
of peak emission in the low resolution Stokes U image.
However, systematic errors make this a
questionable detection with a SNR $\sim$ 2.
As mentioned above,
if this detection is real the signal is
not consistent with a halo because
the polarization would be well
over 100$\%$.  A Stokes I counterpart of GFP emission
has never been seen, but 20 cm GFP
is not well studied.
It is possible that the emission is a new case
of GFP in which the Stokes I has not been completely
resolved out.

\afterpage\clearpage
\vfil\eject

\chapter{Low Resolution Stokes I and the Systematic Errors}

\pagestyle{fancyplain}
{\markright{Chapter 5:~~~Low Resolution Stokes I and the Systematic Errors}
\renewcommand{\chaptermark}[1]%
{\markboth{#1}{}}
\renewcommand{\sectionmark}[1]%
{\markright{Chapter 5:~~~Low Resolution Stokes I and the Systematic Errors}}
\lhead[\fancyplain{}\thepage]%
    {\fancyplain{}\rightmark}
\rhead[\fancyplain{}\leftmark]%
    {\fancyplain{}\thepage}
\cfoot{}


\def\gtwid{\mathrel{\raise.3ex\hbox{$>$\kern-.75em\lower1ex\hbox{$\sim$}}}}
\def\ltwid{\mathrel{\raise.3ex\hbox{$<$\kern-.75em\lower1ex\hbox{$\sim$}}}}


 
We searched for Stokes I halo signals in the low resolution data
after subtracting high resolution CLEAN models of the sources.
The resulting low resolution data appears to be dominated by systematic
errors rather than thermal noise or a halo.  Nonetheless, in the
absence of a clear halo detection, upper limits can be placed on the halo
flux.
This requires a
deeper look into the systematic errors
which we now discuss.
Understanding the systematic errors is also
important to aid others in any future attempts
at this type of work.
The casual
reader may skip to the next section.

One way to investigate the systematic errors is to
look at data
in a case where we know what to expect.  For
example, subtracting CLEAN models of all the sources
in a field from the $uv$ data which produced the image,
should yield zero (neglecting thermal errors).
We refer to this treatment of the data as a
``subtraction'', and to the resulting non-thermal systematic
errors as the ``offsets''.
Our calibrator is a particularly good source
for performing a subtraction because it is
a bright, apparently non-varying point source in a field
remarkably devoid of strong confusing emission.
Unlike the subtractions used to search for halos,
here we subtract data and models from the same
arrays.  This precludes error contributions due to
differently sampled structure.
Figure 5.1 shows the subtraction for the calibrator
1934-638.
One night of data is shown for this object,
observed the first night listed in Table 4.1
for the ``6B Array''.
Each point is the
real part of the
visibility, after subtraction,
for a single antenna pair averaged
over the entire night.
The differences from zero are a clear indication
of systematic errors.
The rms of the offsets is $\sim$ 10 mJy.
We do not understand this error but believe
it originates with the model.
The CLEAN models were made from a wide field self-calibrated image
centered around 1934-638
combining all
9 nights of high resolution data listed in Table 4.1.
The
image has a DR of 100,000:1, and an rms of $\sim$ 150 $\mu$Jy.
Unfortunately for halo searches,
high DR and sensitivity in the image plane is not enough.
As we see in the next section, systematic errors
on the shortest baselines limit the sensitivity
of the search.

\begin{figure}
\begin{center}
\includegraphics[bb=27 170 564 695,width=4.9in]{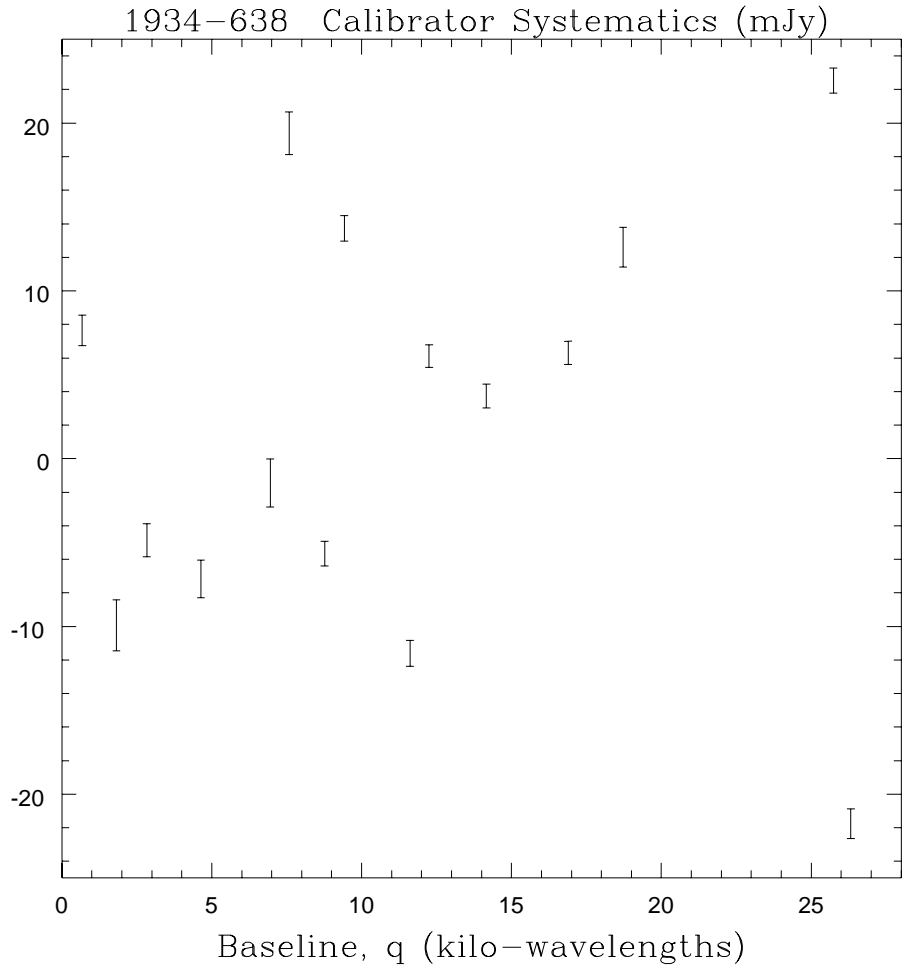}
\end{center}
\caption
{$uv$ data with CLEAN models of the
sources in the field subtracted, for 1934-638.
One night of data is shown for this object taken
on the first night of observations in the ``6B Array''
listed in Table 4.1.
Each point is the
real part of the
visibility, after subtraction,
for a single antenna pair averaged
over the entire night.
The offsets from zero are a clear indication
of systematic errors.}
\label{x4_1934_offsets_1}
\end{figure}

\section{Low Resolution Subtractions}
 
A low resolution subtraction is the difference between
low resolution data and high resolution CLEAN models
of the sources.
This type of subtraction can have errors in addition to
the type seen on the calibrator.  In particular, any
large scale structure unsampled by the high resolution
array will appear as
differences from zero in the low resolution subtraction.
Figure 5.2 shows 5 nights of low resolution subtractions
on 1935-692.
The visibilities for each of the 3 points are
averaged over the entire night.
Only the real part of the complex visibility is shown.
Since we will be assuming radially symmetric models
for a halo, only the real parts of the
visibilities contribute.
The 5 nights shown are for the data collected using
phase switching in the correlator to reduce the
effects of potential DC offsets.
In addition, of the two correlated IF's at 1432 MHz
and 1344 MHz, only the latter is plotted.
We will use these 5 nights of data to
estimate upper limits to the scattered halo flux
and thus upper limts on $\Omega_{\rm IGM}$.
Several features are worth noting.
The data points on individual baselines
are consistent with thermal errors for nights 1 and 3
but are unlikely to be thermal for night 5.
The rms of the offsets from zero
for an individual night is about
1 mJy.
Since the rms of the offsets
for the calibrator in Figure 5.1
is $\sim $ 10 mJy, and the calibrator is 10 times brighter
than 1935-692, the offsets are consistent with
dynamic range
limitations linearly proportional to the peak
source in the field\footnotemark.
\footnotetext{We caution the reader
from concluding that this
$DR_{uv}$ $\sim$ (15 Jy)/(10 mJy) $\sim$ 1,500:1
is the fundamental limit of the telescope,
but instead the limits of our data reduction.}
Unlike the DR defined previously as the ratio of peak
flux to rms in the image, we define $DR_{uv}$
here as the ratio of the peak flux to the
rms, or average offset, of the $uv$ data,
on a particular baseline averaged over time.
This $DR_{uv}$ is most
applicable to our halo search which takes place
in the $uv$ plane.
$DR_{uv}$ is most important for final
halo analysis when evaluated on the shortest baselines.
Systematic errors may result in larger scatter on the
short baselines compared to the rest of the $uv$ plane.
These appear in our data as offsets in the
subtractions as seen in Figure 5.2.
A useful determination of $DR_{uv}$ should include
any contribution from the systematic offsets.
 
 
\begin{figure}
\begin{center}
\includegraphics[bb=27 170 564 695,width=5.4in]{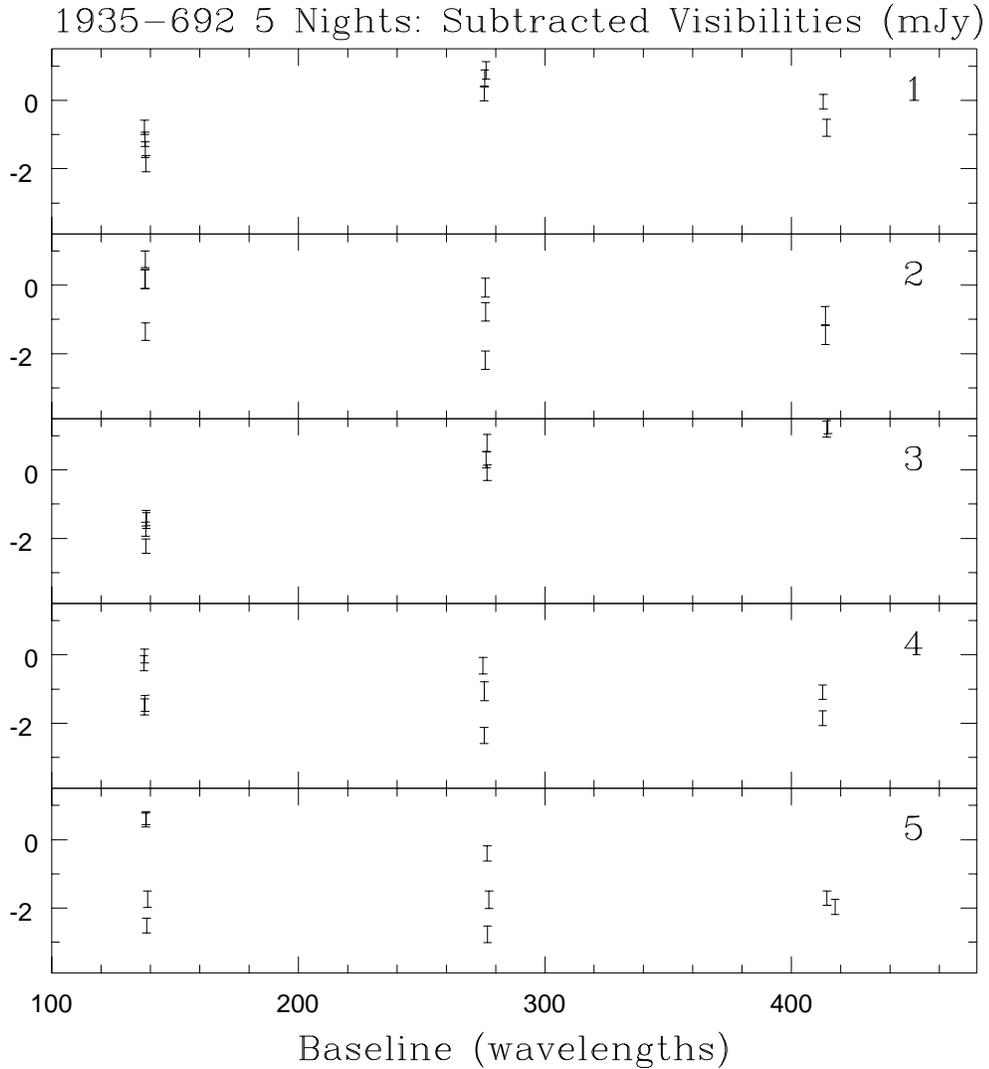}
\end{center}
\caption
{Five nights of low resolution subtractions
on 1935-692.
The visibilities for each of the 3 points are
averages over the entire night.
Only the real part of the complex visibility is shown,
and only for the 1344 MHz IF.
All nights shown are for the data collected using
phase switching in the correlator to reduce the
effects of potential DC offsets.
It is these 5 nights of data which we use to
estimate upper limits to the scattered halo flux,
and thus upper limits on $\Omega_{\rm IGM}$.}
\label{x5_stack_RSTUV_1}
\end{figure}

\begin{figure}
\begin{center}
\includegraphics[bb=27 170 564 695,width=4.9in]{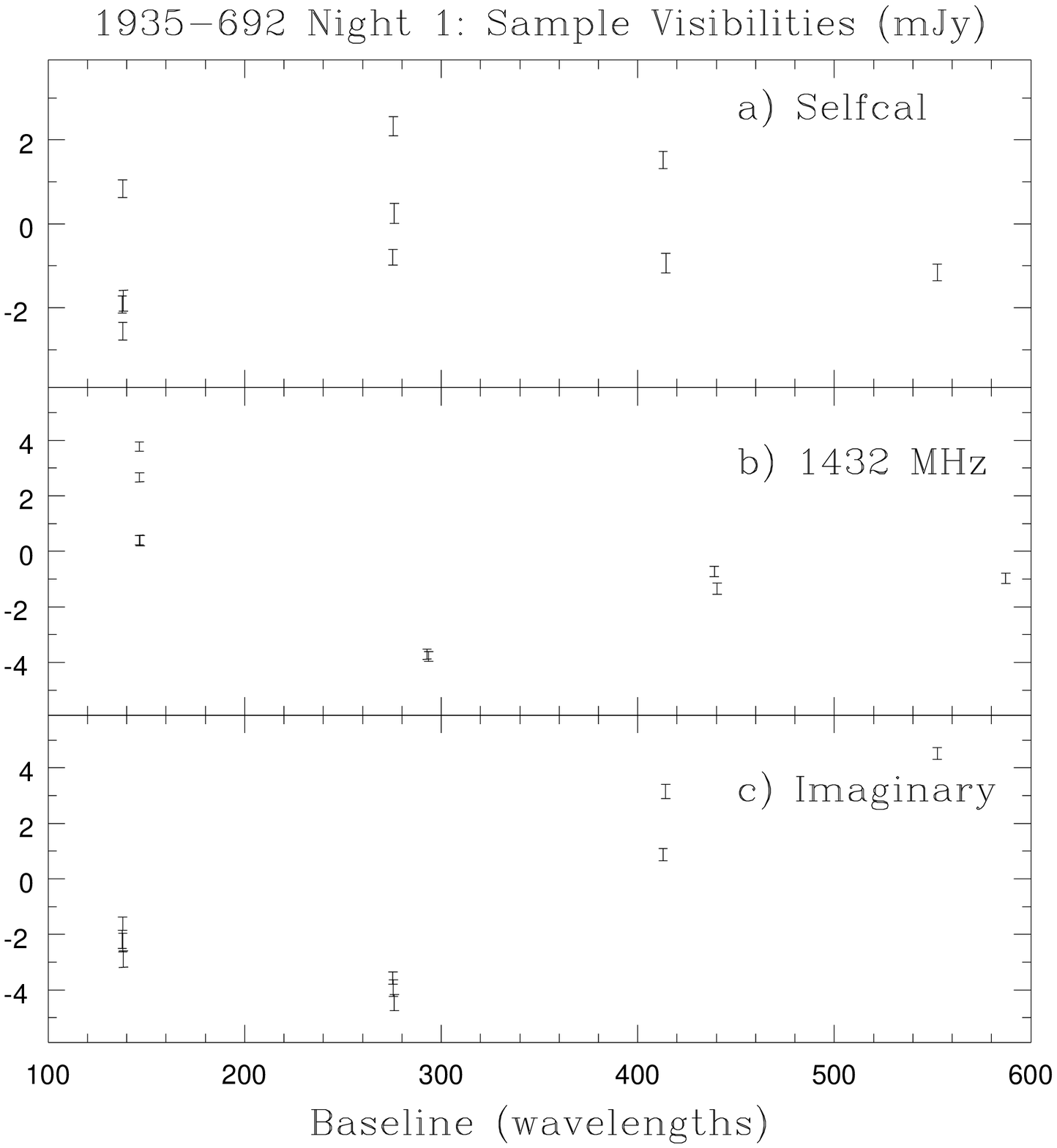}
\end{center}
\caption
{The three plots in this figure are all
visibilities for night 1 of the low resolution array
after subtraction of the high resolution CLEAN models.
This is the same night as the first one shown in
Figure 5.2.  The visibilities for each point are
averages over the entire night.
a) The low resolution data has been self-calibrated
with the high resolution CLEAN model.  Note the
lack of internal consistency,
within thermal errors, between data points
on a given baseline.
b) A low resolution subtraction on the 1432 MHz IF.
c) The imaginary part of the visibility.}  
\label{x6_stack_extras_1}
\end{figure}

The rms in the residual image was our
measure of progress in deciding to move on
from one iteration of Imaging -$>$ Cleaning -$>$ Self-Calibrating
to the next.
In retrospect, we think it is crucial to monitor
the offsets in the $uv$ plane,
or $DR_{uv}$, as well\footnotemark.
\footnotetext{When comparing the $DR_{uv}$ for two
successive iterations of
the usual data reduction cycle, it is important
to use the same model in forming subtractions.
For example,
if three images and data sets are
produced, the first
model should be used to compare the $DR_{uv}$ of
the first and second data sets.
Likewise, the second model should be used to
compare the second and third data sets.
Since monitoring $DR_{uv}$ is meant to track the rms 
of systematic errors
in subtractions, it is important that an improvement
due to a deeper CLEAN is not confused with a decrease
in noise.}  
Although we noticed offsets in the $uv$ plane early
on, we suspected they resulted from faint, uncleaned sources,
and would improve as our image DR improved.
Pursuing this hypothesis was nontrival
and resulted in producing a wide-field image of
1935-692 with an image DR
$\sim $ 77,000:1
(peak/rms), which is 7 times greater than previously obtained at
the ATCA.
In addition, we achieved
a DR $\sim $ 100,000:1 for the calibrator source 1934-638\footnotemark.
 
\footnotetext{
Although these DR's are routinely achieved at telescopes
such as the VLA and WSRT, the ATCA has only
6 antennas (15 baselines).  This
slows down, and potentially reduces, convergence
between data and the true sky brightness distribution.}

Before proceeding to the upper limits, we explain which portions
of the low resolution data were excluded from analysis, and why.
The subtractions for the data with a $\sim$ 1 degree offset
of the phase center resulted in systematic offsets several times
larger than the typical rms of 1 mJy seen in Figure 5.2.
This is most likely due to bandwidth smearing (even though
we attempted to account for that in the data reduction).
We therefore excluded those 5 nights from the halo analysis.
In addition, the observations included two
IF's with a separation of 88 MHz.  Figure 5.3 b) shows a typical
low resolution subtraction on the 1432 MHz IF.  As can be seen,
the magnitude of the offsets are twice as large as those
for the first night shown in
Figure 5.1, which is the same plot for the 1344 MHz IF.
Multi-frequency synthesis
breaks down at large distances from the phase center.
Since we had to CLEAN the entire field of view, we had
no choice but to apply this technique to several distant
sources.  Looking at images
of each IF separately, we have seen that
cleaning resulted in more errors in the 1432 MHz data
than the 1344 MHz data.  We believe this to be the source of
greater offsets in the low resolution subtraction
and have removed the 1432 MHz data from halo analysis.
We also excluded the 122 m baseline from the halo analysis.
Redundancy constraints were applied to the
gain calibration of the redundant spacings of the
122 meter array.
However, the 122 m spacing has no
redundancy and was therefore the least constrained.
As a result, the 122 m baseline had an average offset of
$\sim -3$ mJy.  This is to be compared to the average
offset of $\sim -1$ mJy obtained when only applying
SELFCAL and no redundancy constraints.
In any case, the 122 m spacing has little
leverage in constraining the halo model.
 
Finally, Figure 5.3 c) shows a typical imaginary part
of the low resolution subtraction.
Although the offsets are larger than the real counterparts
in Figure 5.2, they do not directly affect our
particular halo analysis.  This is because we assume
a radially symmetric halo.  For an isotropic
central source, the only way to get a non-radially
symmetric halo is with a lumpy IGM
(e.g. see Katz et. al. (1996)~\cite{katz96}),
or anisotropic illumination.
A lumpy IGM could be significant.
Not only would the halo produce imaginary
parts to the complex visibility, but
higher spatial frequencies could be created by the lumps,
changing the visibility function.

\afterpage\clearpage
\vfil\eject

\chapter{Upper Limits on the IGM Density}

\pagestyle{fancyplain}
{\markright{Chapter 6:~~~Upper Limits on the IGM}
\renewcommand{\chaptermark}[1]%
{\markboth{#1}{}}
\renewcommand{\sectionmark}[1]%
{\markright{Chapter 6:~~~Upper Limits on the IGM}}
\lhead[\fancyplain{}\thepage]%
    {\fancyplain{}\rightmark}
\rhead[\fancyplain{}\leftmark]%
    {\fancyplain{}\thepage}
\cfoot{}


\def\gtwid{\mathrel{\raise.3ex\hbox{$>$\kern-.75em\lower1ex\hbox{$\sim$}}}}
\def\ltwid{\mathrel{\raise.3ex\hbox{$<$\kern-.75em\lower1ex\hbox{$\sim$}}}}


 
Figure 6.1 shows 5 nights of Stokes I low resolution
subtracted data averaged together for each baseline
(These are the 5 nights shown separately in
Figure 5.2).
Each square represents a data point, but the 
thermal error bars have
been left off.
To the right of each set of points, an asterisk
is placed at the
mean value.  The larger error bars around this mean are
the rms of the data points, and the smaller error
bars are the error on the mean
(assumed to be $\sqrt{n}$ smaller).
The three sets of points are for the
31 m, 61 m, and 92 m baselines.  These are the
three shortest baselines in the 122 meter array which we
will use to constrain the halo flux.
Two Stokes I models are shown.
These models use Equation 2.1 for our target source, 1935-692,
with ${\mathcal{S}}_o$ = 1.54 Jy and Z = 3.15.
The ``0.05'' curve is for $\Omega_{\rm IGM}$ = 0.05,
and ``1.00'' is for $\Omega_{\rm IGM}$ = 1.00.
The first feature to note in the data is that each
average is negative.  We think this is the result
of an error in the high resolution model.
Nonetheless, the error in the model is
approximately constant over the three baselines\footnotemark
~(to within $\sim$ 1.5$\sigma_{mean}$).
\footnotetext{It may appear inconsistent that the offsets are
approximately constant over the three baselines shown in
Figure 6.1, yet not constant for the baselines shown
in Figure 5.1 for 1934-638.  
However, the three baseline separations of the low
resolution array in Figure 6.1
are all within 300 wavelengths of each other, as opposed to
the kilo-wavelength separations of Figure 5.1.}
Consequently, we can not measure or constrain
$\Omega_{\rm IGM}$ by measuring the absolute offset
of the real part of the visibility from zero.
But, as can be seen by comparison of
the two models for $\Omega_{\rm IGM}$ = 0.05 and $\Omega_{\rm IGM}$ = 1.00,
$\Omega_{\rm IGM}$ can also be measured or constrained by the
curvature of the sampled visibility function with baseline.

\begin{figure}[p!]
\begin{center}
\includegraphics[bb=27 170 564 650,width=4.9in]{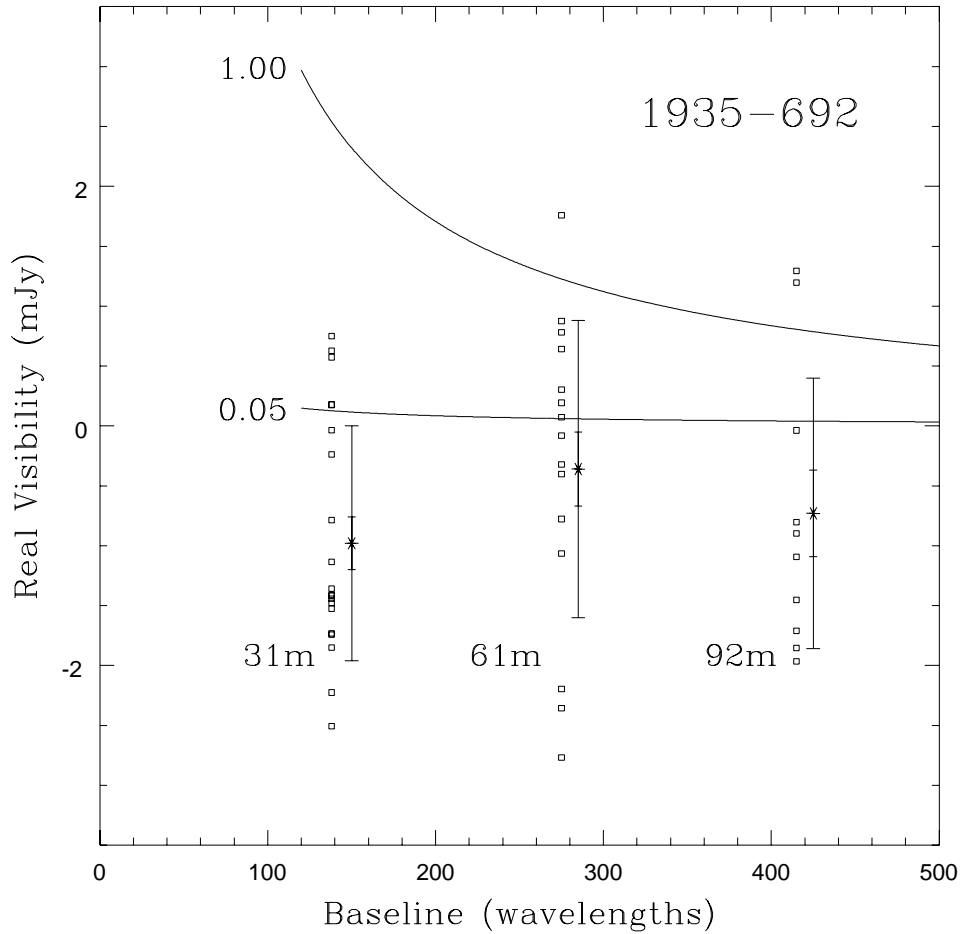}
\end{center}
\caption
{Five nights of Stokes I low resolution
subtracted data averaged together for each baseline
(these are the 5 nights shown separately in
Figure 5.2).
Each square represents a data point, but the thermal error bars have
been left off.
To the right of each set of points is an asterisk at the
mean.  The larger error bars around this mean are
the rms of the data points, and the smaller error
bars are the error on the mean.
The three sets of points are for the
31 m, 61 m, and 92 m baselines.  These are the
three shortest baselines in the 122 meter array, which we
will use to constrain the halo flux.
Two Stokes I models are shown.
These models use Equation 2.1 for our target source, 1935-692,
with ${\mathcal{S}}_o$ = 1.54 Jy and Z = 3.15.
The ``0.05'' curve is for $\Omega_{\rm IGM}$ = 0.05,
and ``1.00'' is for $\Omega_{\rm IGM}$ = 1.00.}
\label{x7_data_mod_7_final_1}
\end{figure}
\afterpage\clearpage
 
\section{ ${\chi}^{2}$ Fit with Two Degrees of Freedom}

Our first method to place an upper limit on
$\Omega_{\rm IGM}$ is a ${\chi}^{2}$ fit with
two degrees of freedom.
We use Equation 2.1 for the model of a halo around
1935-692 (with ${\mathcal{S}}_o$ = 1.54 Jy, Z = 3.15,
$\Omega_o=1.0$, and $H_o =75$).
The first degree of freedom is $\Omega_{\rm IGM}$.
The second degree of freedom is a
parameter called ``shift'', which allows
a global offset for all three baselines.
This is only rigorously appropriate if some
systematic is affecting these three
baselines similarly, which is uncertain.
In this manner, $\Omega_{\rm IGM}$ is constrained only
by the curvature of the measured visibility
function with baseline.
Figure 6.2 shows the ${\chi}^{2}$ contours.
The best fit for the shift is $-0.35$. 
The best fit for $\Omega_{\rm IGM}$ is $-0.25$.
Note that this negative value is not due
to the negative offsets, because those
are accounted for by the shift.
Instead, the negative $\Omega_{\rm IGM}$ is due
to the mean values having a rise between
the 31 m and 61 m baseline in the data, as opposed
to the fall predicted by the model.
Taken literally, the ${\chi}^{2}$ contours indicate a
95$\%$ (or 2$\sigma $) upper limit
of $\Omega_{\rm IGM}$ $<$ 0.15,
but we propose a more conservative approach to the
upper limit.
Suppose the systematic errors on each baseline
had been positive instead of negative.
The magnitude of the systematic errors would
be the same, but the result would be a falling
visibility between the 31 m and 61 m baselines.
For this reflection of the data around zero,
the best fit $\Omega_{\rm IGM}$ would be 0.25,
and the 95$\%$ (or 2$\sigma $)
upper limit would be $\Omega_{\rm IGM}$ $<$ 0.65.
(The ${\chi}^{2}$ contours for this case are just the
contours in Figure 6.2 reflected about the
$\Omega_{\rm IGM}$ origin.)
Therefore, our 2$\sigma $ upper
limit for the ${\chi}^{2}$ fit with a shift is
$\Omega_{\rm IGM}$ $<$ 0.65.
This result assumes an emission age for 1935-692
of $5$x$10^7$ yr, yet is also valid for ages
between $10^7$~yr and $10^8$~yr as discussed in section 3.7.

\begin{figure}[p!]
\begin{center}
\end{center}
\caption
{(This figure has been removed
to reduce size.  It can be obtained
by contacting Robert Antonucci
at ski@ginger.physics.ucsb.edu):
${\chi}^{2}$ contours of the fit with
Equation 2.1 for the model halo visibilities.
A conservative approach to the
upper limit reflects the data around zero and results
in $\Omega_{\rm IGM}$ = 0.25,
and the 95$\%$ (or 2$\sigma $)
upper limit would be $\Omega_{\rm IGM}$ $<$ 0.65.
The ${\chi}^{2}$ contours for this case are just the
contours in the figure reflected about the
$\Omega_{\rm IGM}$ origin.
The contours plotted are for the
probabilities of rejecting the model.
Starting from the inner contour, the
probabilities are:
70$\%$,
80$\%$,
90$\%$,
95$\%$, and
99$\%$.}
\label{x8_chi_contour_final_1}
\end{figure}

\section{${\chi}^{2}$ Fit with One Degree of Freedom and Added Noise}
 
There is another way to account for the systematic
errors without including a global offset (or shift)
as a degree of freedom.
Looking at Figure 6.1, an eyeball estimate of the systematic
offsets is $\sim $ 1 mJy.
Assuming that the true error bars
would reflect this offset,
we add in quadrature, 1 mJy to
the existing errors on the means.
As a result, each mean with its expanded error
bar is consistent with zero, and the significance
of the rise from 31 m to 61 m
is diminished.
Figure 6.3 shows the ${\chi}^{2}$ probabilities vs. $\Omega_{\rm IGM}$
for the model fit with 1 mJy errors added in quadrature.
The model and assumptions are the same
as the previous ${\chi}^{2}$ test with the
exclusion of the shift parameter.
As expected, the best fit $\Omega_{\rm IGM}$ is still negative,
but the 2$\sigma $ upper limit is $\Omega_{\rm IGM}$ $<$ 0.50.

The two ${\chi}^{2}$ tests discussed in this chapter
both assume a gaussian distribution for the rms
of the data points at each baseline.
Since there is no {\it a priori} reason to assume
gaussian errors, a non-parametric test
based on the binomial distribution was performed.
The results are 
similar to the upper limits obtained
with the ${\chi}^{2}$ fits.

\begin{figure}[p!]
\begin{center}
\includegraphics[bb=27 170 564 650,width=4.9in]{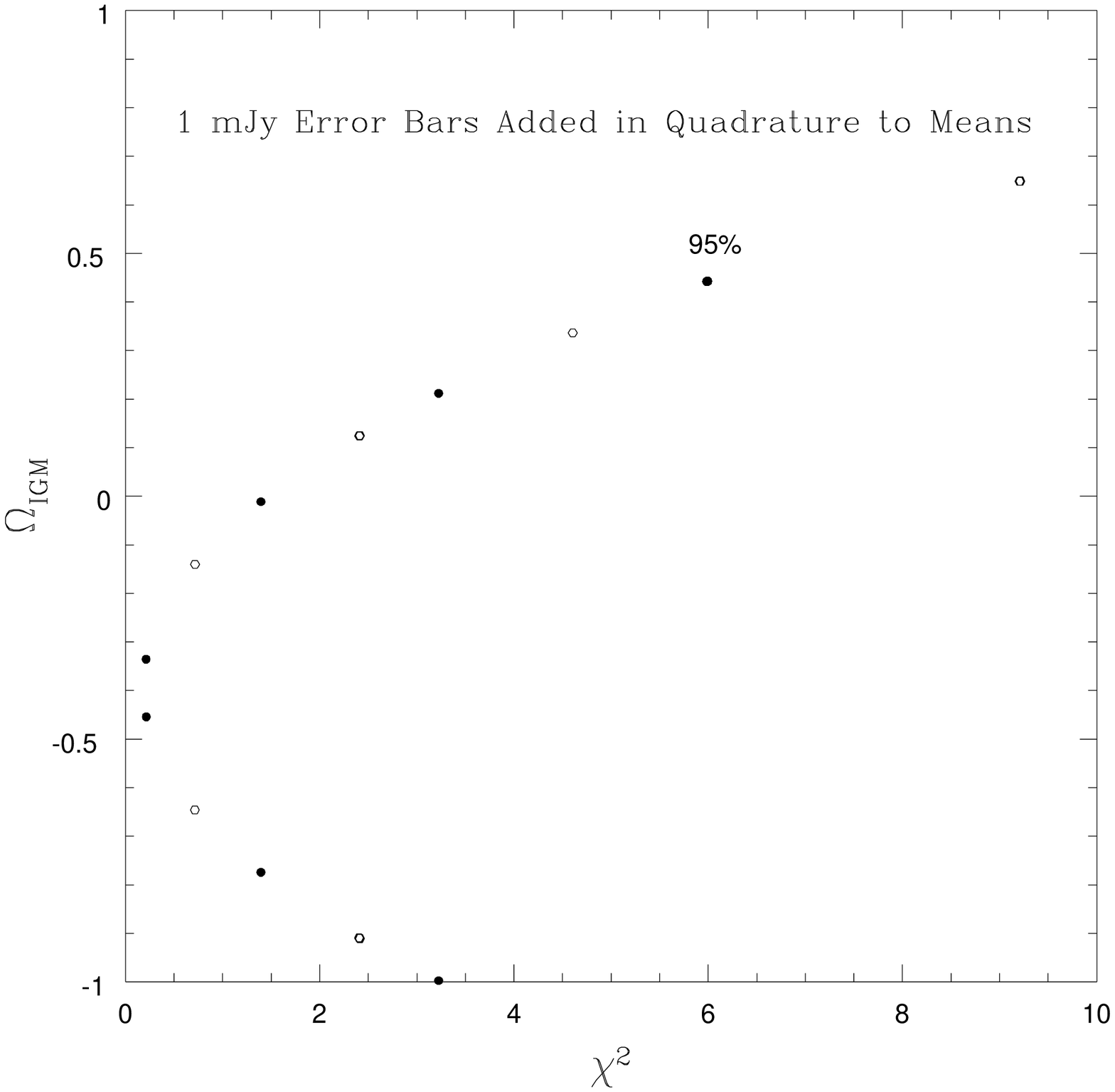}
\end{center}
\caption
{The ${\chi}^{2}$ probabilities vs. $\Omega_{\rm IGM}$
for the model fit with 1 mJy errors added in quadrature.
The best fit $\Omega_{\rm IGM}$ is still negative,
but the 2$\sigma $ upper limit is $\Omega_{\rm IGM}$ $<$ 0.5.
The points plotted are for the
probabilities of rejecting the model.
Starting from the left-most pair of points, the
probabilities are:
10$\%$,
20$\%$,
50$\%$,
70$\%$,
80$\%$,
90$\%$,
95$\%$, and
99$\%$.}
\label{x9_chi_err_final_1}
\end{figure}
\afterpage\clearpage

\chapter{Summary}
 
\pagestyle{fancyplain}
{\markright{Chapter 7:~~~Summary}
\renewcommand{\chaptermark}[1]%
{\markboth{#1}{}}
\renewcommand{\sectionmark}[1]%
{\markright{Chapter 7:~~~Summary}}
\lhead[\fancyplain{}\thepage]%
    {\fancyplain{}\rightmark}
\rhead[\fancyplain{}\leftmark]%
    {\fancyplain{}\thepage}
\cfoot{}


\def\gtwid{\mathrel{\raise.3ex\hbox{$>$\kern-.75em\lower1ex\hbox{$\sim$}}}}
\def\ltwid{\mathrel{\raise.3ex\hbox{$<$\kern-.75em\lower1ex\hbox{$\sim$}}}}

\def\H75{$H_o = 75$}
\def\helium4{$^{4}$He}
\def\He3{$^{3}$He}
\def\Li7{$^{7}$Li}

 
\section{The IGM and Scattered Halos}

Standard big bang nucleosynthesis calculations combined with
observations of the primordial elements 
D, \He3, \helium4, and \Li7
predict an $\Omega_{b}~\sim~3\%$ for \H75.
Only $\sim~10\%$ of these baryons are seen in
stars and hot gas in clusters.
To test the hypothesis that the baryons are
in a diffuse ionized IGM, 
we
searched for Thomson-scattered
halos around strong high redshift radio sources.

Sholomitskii (1990) 
details properties of ``Cosmological Halos'' due
to Thomson scattering of radio waves in the ionized IGM.
The halos are about a degree in size and have an optical
depth $\tau \sim 0.001$
for $\Omega_{\rm IGM}=0.05$, $\Omega_o=1.0$, $H_o =75$,
Z = 3.5, and an assumed emission age of
$5x10^7$ yr.
Therefore a 1 Jy source would have about a 1 mJy
halo, although only 5-10$\%$ of the scattered flux
would be detected at 20 cm for $\sim$ 30 m
baselines of an interferometer.  The halos are
also $\sim$ 1/3 tangentially polarized as
seen in the model for the halo visibility
in Equation 2.1.

To maximize scattered flux, target sources for a
halo search should have strong isotropic emission
and high redshift.
GPS sources probably have more isotropic
emission than larger double sources at high redshift.  But there is no
universally accepted model for these sources,
and they should be used with caution.
It is possible that they are too young
to produce a measurable halo.
Our current search is around the GPS
source 1935-692 which has a 20 cm flux of
1.54 Jy, and Z = 3.15.
 
In general, target sources should be at
$|dec |$ $\gtwid$ $20^{\circ}$
if observed with an East-West array.
This is to provide adequate North-South
resolution for imaging confusing sources in the field.
For polarized halo searches the sources should be $\gtwid$~$30^{\circ}$
away from the Galactic plane. Galactic
Foreground Polarization (GFP) is strongest near the
plane and can completely confuse
the polarization images.  Before lengthly observations are
made, preliminary low resolution images
of candidate fields are necessary to check
against GFP.
 
The strategy to detect halos requires low resolution
baselines to sample the halo, and high resolution
baselines to image the confusing sources.
At 20 cm, $\sim$ 30 m spacings can sample 20
arcmin structure of the halo, while spacings out
to $\sim$ 6 km can make high ($\sim$ 5 arcsec)
resolution images of confusing sources.
Models of confusing sources can be made
using CLEAN algorithms and subtracted from the
low resolution $uv$ data.  Ideally, this
``subtraction'' results in low resolution
data containing only the response of a faint halo.
However, large scale confusing emission and
systematic errors limit the sensitivity of
the search.

\section{Observations and Results}
 
We observed 1935-692 at the ATCA for 9 nights
in high resolution arrays and 10 nights in a
low resolution array (see Table 4.1).
We also interleaved a secondary
calibrator, 1934-638, with a duty cycle
chosen to result in a dynamic range (DR) comparable to 
that of the target
source.  We found the calibrator observation
to be invaluable for investigating
systematic errors, and highly recommend
its inclusion in any observing program.
Redundancy of the shortest spacings
in the low resolution array
was used to help calibrate 
antenna gains.

High resolution images with high DR were made for
1935-692 and 1934-638.  These were used to make CLEAN
models of sources for both calibration of,
and subtraction from, the low resolution data.
We achieved DR's
(peak/rms) of 77,000:1 and 100,000:1,
for 1935-692 and 1934-638 respectively.
These are significant increases over
the previously obtained DR's at the ATCA of
around 10,000:1.
 
Low resolution polarization imaging of 1935-692
revealed large scale polarized emission in both
Stokes Q and U.  The emission is not
centered around 1935-692
and is more consistent with GFP than with a
cosmological halo.  GFP is characterized
by having little or no detectable Stokes I emission,
resulting in apparent polarization fractions $>>$ 100$\%$.
Halo models unequivocally predict detected polarization
fractions $<100\%$.
Internal consistency and checks with the
calibrator
lead us to
believe the the emission is real.
While it is not absolutely
certain that the signal is
truly GFP, the emission prevents us
from placing meaningful limits on polarized
scattered halo flux.
 
The Stokes I analysis for halos was limited by
systematic errors.
We obtained high DR in the
image plane. However, it is the $DR_{uv}$
in the $uv$ plane on the shortest baselines averaged over time
which is most important.
The $DR_{uv}$ is the ratio of the peak flux to the rms, or
average offset, 
on a particular baseline averaged over some time interval.
For 1935-692, we had an rms $\sim$ 1 mJy
for the offsets on an individual
night.  With a peak flux of
${\mathcal{S}}_o$ = 1.54 Jy,
this  is a $DR_{uv} \sim$ 1,500:1 on the
shortest baselines in the 122 meter array.
Hindsight has shown that monitoring the $DR_{uv}$
as a measure of progress is more meaningful than
the image plane DR since the halo analysis takes place
in the $uv$ plane.
We recommend monitoring $DR_{uv}$
on both target and calibrator source, and on
high and low resolution subtractions (if multiple
arrays are used).
 
We think our systematic errors
derive ultimately from imperfections in the CLEAN
models of the sources, but do not understand the origin.
The systematic errors affected different segments of our data
differently, and we excluded from the halo analysis
those segments with largest errors.
One excluded segment is 5 nights of data taken
with a $\sim$ 1 degree phase offset from the pointing center.
The other segment excluded is one of the two IF's
for which the multi-frequency synthesis produced
a poorer model.
 
Figure 6.1 shows the low resolution subtracted
data used for halo analysis.  Each baseline has a
negative offset which we think is due to a
systematic error, as opposed to a
difference indicative of real emission on
the sky.
The systematics prevent any clear detection
or tight limit on
a halo.  Nonetheless, an upper limit can
be placed on any scattered halo flux
and thus on $\Omega_{\rm IGM}$.
Our first method to determine an upper limit
is a ${\chi}^{2}$ fit
to Equation 2.1 with 2 degrees of freedom.
In this fit, one free parameter is $\Omega_{\rm IGM}$.
The other free parameter is a constant ``shift'',
included to account for the systematic, negative
offset of the data.
This allows a fit of $\Omega_{\rm IGM}$ through the
relative changes of the baselines.  Our
upper limit based on this
method is $\Omega_{\rm IGM} < 0.65 ~ (2\sigma, 95\%)$.

There is another way to account for the systematic
errors without assuming a global offset (or shift).
We can accomodate the offset by adding
1 mJy in quadrature to
the existing errors on the means for each baseline.
As a result, each mean with its expanded error
bar is consistent with zero, and the significance
of the rise 
(probably unphysical) from 31 m to 61 m
is diminished.
With this ${\chi}^{2}$ fit the 2$\sigma $
upper limit is $\Omega_{\rm IGM}$ $<$ 0.50.
 

\afterpage\clearpage
\vfil\eject

\chapter{Conclusion}
 
\pagestyle{fancyplain}
{\markright{Chapter 8:~~~Conclusion}
\renewcommand{\chaptermark}[1]%
{\markboth{#1}{}}
\renewcommand{\sectionmark}[1]%
{\markright{Chapter 8:~~~Conclusion}}
\lhead[\fancyplain{}\thepage]%
    {\fancyplain{}\rightmark}
\rhead[\fancyplain{}\leftmark]%
    {\fancyplain{}\thepage}
\cfoot{}


\def\gtwid{\mathrel{\raise.3ex\hbox{$>$\kern-.75em\lower1ex\hbox{$\sim$}}}}
\def\ltwid{\mathrel{\raise.3ex\hbox{$<$\kern-.75em\lower1ex\hbox{$\sim$}}}}



We would like the reader to see that the search for
halos has motivations in fundamental cosmology and
that this current attempt finds $\Omega_{\rm IGM}<1$.
In addition, this thesis is meant to provide helpful
hints for others wishing to search for cosmological
halos at the ATCA, WSRT, or the VLA.

Although the halo project is very difficult, it is
robust against false detections.  An observer can
have confidence in a detection if
1) The radial profile follows Equation 2.1;
2) There is a corresponding low signal on the imaginary part
of the visibilities;
3) The predicted magnitude and direction of polarization
is seen; and 4) A low redshift calibrator
produces {\it no} halo since the density of ionized gas
at low Z greatly decreases the optical depth.
In addition, the ${\mathcal{S}}_{o}$ and Z dependence
in Equation 2.1
can be used to confirm
$\Omega_{\rm IGM}$ in a survey of different objects.
Furthermore, Equation 2.1 can be used to coadd
multiple halo visibilities of the same, or different, sources
observed on any telescope arrays.
Our experience has been that both systematic errors
in Stokes I and large size-scale polarized confusion
can prevent reaching thermal sensitivity on the
short baselines where it is needed.
Therefore, we suggest that future observing programs
involve several halo targets, and
accepting sources with only
modest thermal SNR estimates.
Coadding the results of the survey makes full use of all
healthy data, minimizes superfluous integration time,
and hopefully averages down systematic errors.
Finally, with the increased sensitivity becoming
available because of receiver improvements, it becomes
affordable to target extended doubles rather
than GPS sources.
Though fainter, the extended sources should be
safer regarding assumptions about isotropic emission and
ages $\gtwid 10^{7}$ years.

\afterpage\clearpage
\vfil\eject


\bibliographystyle{nature}
\bibliography{thesis}

\end{document}